\documentclass[12pt]{article}
\usepackage{caption}
\usepackage{amssymb}
\usepackage{amsmath}
\usepackage{hyperref}
\usepackage{graphicx}
\usepackage{float}
\usepackage{bm}
\usepackage{latexsym}
\usepackage{subfig}
\usepackage{mathtools}
\usepackage{booktabs}
\usepackage{tabularx}   
\usepackage{geometry} 
\usepackage{placeins}
\usepackage{subfig}
\usepackage{subcaption}
\begin{document}

\title{\bf Topological Classification of a $4D$ AdS Black Hole with Non-Minimal Maxwell Coupling}

\author{
Faramarz Rahmani\thanks{Corresponding author: faramarz.rahmani@abru.ac.ir}
\and
Mehdi Sadeghi\thanks{Email: mehdi.sadeghi@abru.ac.ir}
}

\date{\today}

\maketitle

\begin{center}
{\small
Department of Physics, Faculty of Basic Sciences,\\
Ayatollah Boroujerdi University, Boroujerd, Iran
}
\end{center}

\abstract{
We perform a topological classification of the phase structure of a four-dimensional AdS black hole with non-minimal Maxwell coupling. Critical points are treated as topological defects, allowing us to assign a winding number to each black hole branch and compute the global topological invariant \(W\).
The system exhibits a duality governed by its Maxwell charge \(Q\): for large \(Q\) it falls into the class \(W = +1\), displaying van der Waals–type behavior with a first-order small/large black hole transition. For small \(Q\), it shifts to \(W = 0\), characteristic of a Hawking–Page transition. This topological classification provides a model-independent validation of the conventional thermodynamic analysis.
Crucially, we find that the non-minimal coupling \(\lambda\) stabilizes the Hawking–Page universality class (\(W=0\)) for black holes with non-zero charge, a phenomenon absent in the standard Reissner–Nordström–AdS case. This establishes a direct link between the microscopic coupling and the macroscopic topological class, demonstrating the power of topological methods in decoding thermodynamic universality across modified gravity theories.
}

\noindent PACS numbers: 11.25.Tq, 04.70.Dy, 04.50.Kd, 04.20.-q , 04.70.bw,05.70.Ce\\

\noindent \textbf{Keywords:} Black hole phase transition, AdS black holes, Topological classification of black holes, Generalized free energy.

\section{Introduction}\label{intro}

The transition from Newtonian mechanics to the theories of special and general relativity constituted a profound paradigm shift in our understanding of spacetime and gravitation. In parallel, the development of quantum mechanics despite its extraordinary empirical success, revealed an intrinsically probabilistic description of nature whose conceptual foundations remain deeply debated.

The synthesis of relativity and quantum mechanics stands as one of the central achievements of twentieth-century physics. Dirac’s formulation of the relativistic wave equation for the electron marked an essential early milestone~\cite{Dirac1927}, while the subsequent development of quantum field theory (QFT) was shaped by many foundational contributions. These include the quantization of the electromagnetic field~\cite{Dirac1928}, the formulation of renormalization theory by Tomonaga, Schwinger, and Feynman~\cite{Tomonaga1946,Schwinger1948,Feynman1949}, and the construction of the electroweak theory by Glashow, Salam, and Weinberg~\cite{Glashow1961,Salam1968,Weinberg1967}. The proof of renormalizability for non-Abelian gauge theories by ’t~Hooft and Veltman~\cite{tHooft1971,tHooft1972} ultimately established the mathematical consistency of the Standard Model.

Within this framework, the fundamental entities are dynamical fields, with particles emerging as their quantized excitations. Interactions are encoded through local terms in the Lagrangian density, and the resulting quantum theory provides an exceptionally accurate description of all known non-gravitational forces. This success has reinforced a guiding principle: a complete and consistent description of nature should admit a quantum formulation of all interactions, including gravity. This conviction underlies the long-standing quest for a theory of quantum gravity.

Attempts to quantize gravity, however, have encountered formidable obstacles. Early efforts to apply perturbative QFT techniques to general relativity, treating the metric as a massless spin-2 field, revealed the presence of uncontrollable ultraviolet divergences. The resulting theory is non-renormalizable, in the sense that divergences cannot be absorbed into a finite number of physical parameters~\cite{Deser1974,tHooft1974,Veltman1975,Goroff1986}. This strongly suggests that general relativity should be regarded as an effective field theory, valid at low energies but requiring an ultraviolet-complete description at the Planck scale.

This impasse has motivated the development of approaches that go beyond the standard perturbative QFT paradigm. Two of the most prominent are string theory and loop quantum gravity.  

String theory~\cite{Green1987,Polchinski1998} postulates that the fundamental degrees of freedom are one-dimensional extended objects whose vibrational modes correspond to particle states, including a massless spin-2 excitation identified with the graviton~\cite{Scherk1974}. The extended nature of strings softens ultraviolet divergences, offering a framework in which gravity and gauge interactions may be unified within a finite quantum theory~\cite{Maldacena1999}.  

By contrast, loop quantum gravity (LQG)~\cite{Rovelli2004,Ashtekar2004,Thiemann2007} pursues a non-perturbative canonical quantization of general relativity that preserves its geometric foundations. Formulated in terms of Ashtekar connection variables, LQG predicts a discrete spectrum for geometric observables such as area and volume~\cite{Ashtekar1997,Rovelli1995}, replacing the classical spacetime continuum with a Planck-scale spin-network structure. One of its notable achievements is a microscopic derivation of the Bekenstein--Hawking black hole entropy~\cite{Rovelli1996,Ashtekar2000}.

Complementary to these fundamental approaches, the study of higher-order and modified theories of gravity provides a pragmatic avenue for probing quantum gravitational effects. Motivated in part by the demonstration that general relativity is perturbatively non-renormalizable~\cite{Boulware1985}, higher-curvature extensions were proposed to restore renormalizability~\cite{Stelle1977}. Although such theories often suffer from ghost-like instabilities, they play a crucial role in establishing gravity as an effective quantum field theory.

A major breakthrough was the realization that special combinations of higher-curvature terms, most notably Gauss-Bonnet and more generally Lovelock gravities~\cite{Lovelock1971}, lead to second-order field equations and can evade ghost pathologies. These theories arise naturally in the low-energy effective action of string theory~\cite{Gross1986}, thereby providing a concrete link between fundamental quantum gravity proposals and phenomenological modifications of general relativity.

Over the past decades, this program has expanded to encompass a broad class of modified gravity theories, including $f(R)$ gravity~\cite{Sotiriou2010}, Horndeski theory~\cite{Horndeski1974,Kobayashi2011}, and Einsteinian cubic gravity~\cite{Bueno2016}. Such models serve as controlled laboratories for exploring phenomena that remain inaccessible in a complete theory of quantum gravity, including the resolution of spacetime singularities~\cite{Charmousis2012}, the microscopic origin of black hole entropy~\cite{Myers2009}, and novel phase structures in black hole thermodynamics~\cite{Cai2013}.

Beyond technical constructions, deep conceptual questions challenge the very premises of the problem. A notable dissenting perspective, advocated by Penrose~\cite{Penrose2004,Penrose2016}, argues that the difficulty in quantizing gravity may signal an incompleteness in quantum mechanics itself, particularly in relation to the measurement problem. In this view, wavefunction collapse is conjectured to be an intrinsically gravitational phenomenon, suggesting that unification may require a revision of quantum foundations rather than a direct quantization of gravity.

In a related but distinct direction, the thermodynamic properties of black holes have motivated the idea that gravity may be an emergent phenomenon. The laws of black hole mechanics~\cite{Bardeen1973} and the discovery of Hawking radiation~\cite{Hawking1975} point to a deep interplay between gravity, thermodynamics, and quantum theory. This insight culminated in the derivation of Einstein’s equations from thermodynamic arguments~\cite{Jacobson1995} and inspired the broader paradigm of emergent gravity~\cite{Padmanabhan2010,Verlinde2011}, in which spacetime geometry arises as a macroscopic manifestation of underlying microscopic degrees of freedom.

The thermodynamics of AdS black holes has received renewed attention through the AdS/CFT correspondence~\cite{Maldacena1999,Gubser1998,Witten1998}, which relates gravitational dynamics in anti-de Sitter space to a conformal field theory defined on its boundary. Within this framework, the Hawking--Page phase transition between thermal AdS and AdS black holes~\cite{Hawking1983} admits a dual interpretation as a confinement--deconfinement transition in the boundary gauge theory~\cite{Witten1998}.  

This correspondence has revealed a rich phenomenology in black hole thermodynamics. In the extended phase space, where the cosmological constant is promoted to a thermodynamic pressure~\cite{dolan2011,Kastor2009,Kubiznak2012}, AdS black holes exhibit phase behavior strikingly analogous to that of ordinary thermodynamic systems. Van der Waals–type phase transitions~\cite{Kubiznak2012,hendi2018}, reentrant phase transitions~\cite{Altamirano2013}, and triple points~\cite{Altamirano2014} have been identified, establishing black holes as valuable analog systems for studying critical phenomena.

More recently, topological methods have been introduced to provide a global and coordinate-independent classification of black hole phase transitions~\cite{Wei:2022dzw,Wei:2024gfz,Wei2022,Yerra2022}. In this approach, thermodynamic critical points are interpreted as topological defects in parameter space. By constructing an appropriate thermodynamic potential and employing Duan’s $\phi$-mapping theory, a topological charge (or winding number) can be assigned to each critical point. The sum of these charges defines a global topological invariant that characterizes the phase structure of the system, distinguishing universal classes such as standard Van der Waals behavior, from more exotic phase patterns independently of the microscopic gravitational dynamics. Parallel developments involving holographic entanglement entropy~\cite{Ryu2006,Nishioka2009} and complexity~\cite{Stanford2014,Brown2016} have further enriched this picture by linking thermodynamic phases to quantum information measures in the dual field theory.

Our research program has contributed to this line of investigation through a series of studies on phase transitions in extended gravitational frameworks. In particular, we have analyzed the thermodynamic and holographic properties of non-minimal Yang--Mills AdS black branes~\cite{Sadeghi:2023tzf,Sadeghi2023a,Sadeghi2023b}, examined the phase structure of four-dimensional Gauss--Bonnet black holes with Yang--Mills charge in a cloud of strings~\cite{Rahmani2024}, and performed a detailed conventional thermodynamic analysis of the four-dimensional non-minimal Maxwell--AdS black hole considered in the present work~\cite{Sadeghi2025}.

The aim of this paper is to extend our previous analysis~\cite{Sadeghi2025} by applying the topological classification framework to the non-minimal Maxwell--AdS black hole. By computing the topological charges associated with its thermodynamic critical points, we provide a global and model-independent characterization of the phase structure, which interpolates between Hawking--Page–type and Van der Waals–type behavior. This synthesis of detailed thermodynamic analysis with topological methods not only sharpens the understanding of the present system but also contributes to the broader program of employing black hole thermodynamics as a probe of the fundamental nature of gravity.

\section{Conventional Thermodynamic Analysis: Essential Results}
\label{sec2}

Nonlinear models play a central role in the study of gravitational systems, as they provide a natural framework for capturing the intricate interactions underlying black hole thermodynamics. The analytical solution of the model considered here was obtained in our previous work~\cite{Sadeghi2025} by means of a perturbative approach. To properly contextualize the topological analysis developed in the present study, it is useful to briefly recall the essential features of that solution. Accordingly, this section summarizes the key thermodynamic quantities, namely the Hawking temperature, entropy, and Gibbs free energy without repeating the technical details of the derivations. Our primary objective is to delineate the conventional phase structure of the system, which will subsequently be reinterpreted and globally classified using topological methods in Sec.~\ref{sec3}.

\subsection{The Action and Solutions}

The model analyzed in Ref.~\cite{Sadeghi2025} is governed by the action
\begin{align}\label{action}
S = \int d^{4} x \sqrt{-g} \bigg[ & \frac{1}{2}(R - 2\Lambda) - \frac{1}{4} F_{\mu \nu} F^{\mu \nu} + \nonumber \\
& \frac{\lambda}{2} F_{\alpha \beta} F_{\mu \nu} \left( g^{\nu \beta} R^{\mu \alpha} - g^{\nu \alpha} R^{\mu \beta} - g^{\mu \beta} R^{\nu \alpha} + g^{\mu \alpha} R^{\nu \beta} \right) \bigg],
\end{align}
where $R$ denotes the Ricci scalar, $\Lambda=-3/l^{2}$ is the negative cosmological constant associated with the AdS radius $l$, and $F_{\mu\nu}=\partial_{\mu}A_{\nu}-\partial_{\nu}A_{\mu}$ is the Maxwell field strength constructed from the gauge potential $A_{\mu}=(h(r),0,0,0)$. The coupling parameter $\lambda$, with dimension $[\lambda]=[L]^2$, characterizes the non-minimal interaction between the electromagnetic field and spacetime curvature~\cite{Balakin:2005fu}. The physical motivation and consistency of this action have been discussed in detail in Ref.~\cite{Sadeghi2025}.

As the system does not admit an exact closed-form solution, we proceed perturbatively. Since the coupling constant $\lambda$ is dimensionful, its smallness is naturally defined relative to the AdS curvature scale $l$. We therefore adopt the static, spherically symmetric metric ansatz
\begin{equation}\label{metric}
ds^{2} = - f(r)e^{-2H(r)} dt^{2} + \frac{dr^{2}}{f(r)} + r^{2} (d\theta^{2} + \sin^{2}\theta\, d\phi^{2}),
\end{equation}
where the function $H(r)$ encodes the deviations induced by the non-minimal coupling.

Varying the action \eqref{action} with respect to the metric $g_{\mu\nu}$ leads to the modified Einstein equations
\begin{equation}\label{EOM1}
R_{\mu \nu} - \frac{1}{2} g_{\mu \nu} R + \Lambda g_{\mu \nu} = T^{\text{(eff)}}_{\mu \nu},
\end{equation}
with an effective energy--momentum tensor given by
$T^{\text{(eff)}}_{\mu \nu} = T^{\text{(M)}}_{\mu \nu} + \lambda T^{(I)}_{\mu \nu}$.
Here, $T^{\text{(M)}}_{\mu \nu}$ is the standard Maxwell energy--momentum tensor,
\begin{equation}
T^{\text{(M)}}_{\mu \nu} = F_{\mu}{}^{\alpha} F_{\nu \alpha} - \frac{1}{4} g_{\mu \nu} F_{\alpha \beta} F^{\alpha \beta},
\end{equation}
while $T^{(I)}_{\mu \nu}$ encodes the contribution arising from the non-minimal interaction. Its explicit form reads
\begin{align}\label{tt}
-2 T^{(I)}_{\mu \nu } &= 2 F_{\mu }{}^{\alpha } F_{\nu }{}^{\beta } R_{\alpha \beta } -  F_{\alpha }{}^{\gamma } F^{\alpha \beta } g_{\mu \nu } R_{\beta \gamma } - 2 F_{\alpha }{}^{\beta } F_{\nu }{}^{\alpha } R_{\mu \beta } \nonumber \\ 
&\quad - 2 F_{\alpha }{}^{\beta } F_{\mu }{}^{\alpha } R_{\nu \beta } + F_{\nu }{}^{\alpha } \nabla_{\beta }\nabla^{\beta }F_{\mu \alpha } + F_{\mu }{}^{\alpha } \nabla_{\beta }\nabla^{\beta }F_{\nu \alpha } \nonumber \\ 
&\quad + F^{\alpha \beta } g_{\mu \nu } \nabla_{\beta }\nabla_{\gamma }F_{\alpha }{}^{\gamma } + F_{\nu }{}^{\alpha } \nabla_{\beta }\nabla_{\mu }F_{\alpha }{}^{\beta } + F^{\alpha \beta } \nabla_{\beta }\nabla_{\mu }F_{\nu \alpha } \nonumber \\ 
&\quad + F_{\mu }{}^{\alpha } \nabla_{\beta }\nabla_{\nu }F_{\alpha }{}^{\beta } + F^{\alpha \beta } \nabla_{\beta }\nabla_{\nu }F_{\mu \alpha } + 2 \nabla_{\beta }F_{\mu }{}^{\alpha } \nabla^{\beta }F_{\nu \alpha } \nonumber \\ 
&\quad + g_{\mu \nu } \nabla_{\beta }F^{\alpha \gamma } \nabla_{\gamma }F_{\alpha }{}^{\beta } -  g_{\mu \nu } \nabla_{\alpha }F^{\alpha \beta } \nabla_{\gamma }F_{\beta }{}^{\gamma } \nonumber \\ 
&\quad + F^{\alpha \beta } g_{\mu \nu } \nabla_{\gamma }\nabla_{\beta}F_{\alpha }{}^{\gamma } + \nabla_{\beta }F_{\nu }{}^{\alpha } \nabla_{\mu }F_{\alpha }{}^{\beta } + \nabla_{\beta }F^{\alpha \beta } \nabla_{\mu }F_{\nu \alpha } \nonumber \\ 
&\quad + \nabla_{\beta }F_{\mu }{}^{\alpha } \nabla_{\nu }F_{\alpha }{}^{\beta } + \nabla_{\beta }F^{\alpha \beta } \nabla_{\nu }F_{\mu \alpha } .
\end{align}

Variation of the action \eqref{action} with respect to the gauge potential $A_{\mu}$ yields the modified Maxwell equations,
\begin{equation}\label{EOM-YM}
\nabla_{\nu}\left( -\frac{1}{2} F^{\mu \nu } -2 \lambda F^{\nu \alpha } R^{\mu }{}_{\alpha } +2 \lambda F^{\mu \alpha } R^{\nu }{}_{\alpha }\right) = 0.
\end{equation}

To solve the coupled field equations, we expand the metric functions and gauge potential perturbatively up to first order in $\lambda$,
\begin{align}
f(r) &= f_0(r) + \lambda f_1(r), \label{f} \\
H(r) &= H_0(r) + \lambda H_1(r), \label{H} \\
h(r) &= h_0(r) + \lambda h_1(r). \label{h}
\end{align}
Substituting these expansions into the equations of motion and solving order by order, while imposing regularity at the perturbed horizon and fixing the integration constants, yields the following first-order solutions~(see Ref.~\cite{Sadeghi2025} for details):
\begin{equation}
H(r) = -\frac{6 Q^2}{r^4}\lambda,
\end{equation}
\begin{equation}\label{hnew}
h(r) = \left(\frac{1}{r} - \frac{1}{r_h}\right) Q + \left[\left(\frac{8 \Lambda}{r} - \frac{8 \Lambda}{r_h}\right) Q + \left(\frac{2}{5 r^{5}} - \frac{2}{5 r_h^{5}}\right) Q^{3}\right] \lambda,
\end{equation}
and
\begin{equation}
\begin{aligned}
f(r) =\ & 1 - \frac{\Lambda r^2}{3} 
+ \left( \frac{r_h^3 \Lambda}{3} - \frac{Q^2}{2r_h} - r_h \right) \frac{1}{r} 
+ \frac{Q^2}{2 r^2} \\
& + \lambda \left[ \left( -\frac{2 Q^2 \Lambda}{r_h} - \frac{2 Q^4}{5 r_h^5} + \frac{2 Q^2}{r_h^3} \right) \frac{1}{r} + \frac{16 Q^2 \Lambda}{3 r^2} - \frac{12 Q^2}{r^4} \right. \\
& \quad \left. + \left( - \frac{10 Q^2 r_h^3 \Lambda}{3} + \frac{5 Q^4}{r_h} + 10 Q^2 r_h \right) \frac{1}{r^5} - \frac{23 Q^4}{5 r^6} \right].
\end{aligned}
\end{equation}

To determine the conserved mass, we expand the time--time component of the metric, $g_{tt}=e^{-2H(r)}f(r)$, in powers of $\lambda$ and compare it with the standard asymptotic AdS form,
\begin{equation}
g_{tt} = 1 - \frac{\Lambda r^2}{3} - \frac{2m}{r} + \cdots .
\end{equation}
This procedure yields the ADM mass of the black hole to first order in $\lambda$,
\begin{equation}\label{mass}
m = \frac{r_h}{2} - \frac{r_h^3 \Lambda}{6} + \frac{Q^2}{4 r_h} + \left( \frac{Q^2 \Lambda}{r_h} + \frac{Q^4}{5 r_h^5} - \frac{Q^2}{r_h^3} \right) \lambda
= m_0 + m_1 \lambda .
\end{equation}

A potential subtlety arises from the presence of higher-order curvature couplings in AdS spacetime, where a holographic renormalization procedure is sometimes required to define conserved charges. In the present case, however, it was shown in Ref.~\cite{Sadeghi2025} that the ADM and renormalized holographic methods lead to identical results. This equivalence follows from the fact that the perturbative corrections proportional to $\lambda$ modify only the value of the mass parameter without altering the asymptotic falloff structure of the solution; consequently, the mass information remains fully encoded in the coefficient of the $1/r$ term. Having established the gravitational solution and its conserved mass, we now proceed to summarize the thermodynamic quantities relevant for the subsequent topological analysis of the phase structure.

\subsection{Summary of Thermodynamic Results}
\label{subsec:thermodynamics}
We summarize the essential thermodynamic properties of the non-minimally coupled black hole solution below. These results form the foundational basis for our subsequent topological investigation.

The fundamental thermodynamic quantities, accurate to first order in the coupling constant \(\lambda\), are the enthalpy (identified with the ADM mass), temperature, and entropy \cite{Sadeghi2025}:

\begin{equation}
\label{enthalpy}
\mathcal{H} = \frac{r_h}{2} + \frac{4 \pi P r_h^3}{3} + \frac{Q^2}{4 r_h} + \left(-\frac{8 \pi P Q^2}{r_h} + \frac{Q^4}{5 r_h^5} - \frac{Q^2}{r_h^3} \right) \lambda,
\end{equation}

\begin{equation}
\label{temperature}
T = \frac{1}{2 \pi} \left[ \frac{1}{\sqrt{g_{rr}}} \frac{d}{dr} \sqrt{-g_{tt}} \right] \Bigg|_{r = r_h} = \frac{1}{4\pi r_h}+ 2 P r_h - \frac{Q^2}{8\pi r_h^3} + \left( -\frac{4 Q^2 P}{r_h^3} + \frac{Q^2}{2\pi r_h^5} \right) \lambda,
\end{equation}

\begin{equation}
\label{entropy}
S = \int \frac{1}{T} \frac{d\mathcal{H}}{dr_h} \, dr_h = \pi r_h^2 - \frac{4\pi Q^2 \lambda}{r_h^2}.
\end{equation}

The system exhibits a rich phase structure that interpolates between Hawking-Page and van der Waals-type transitions. A conventional thermodynamic analysis reveals that increasing the Maxwell charge \(Q\) promotes van der Waals-like behavior, while decreasing it favors a Hawking-Page-type phase transition. In the former regime, the canonical ensemble displays swallowtail structures in the free-energy–temperature (\(F\)-\(T\)) diagrams. In the latter regime, the diagrams indicate a transition between thermal AdS space and a large black hole branch at the Hawking-Page temperature.

Furthermore, for small charge, the temperature–horizon-radius (\(T\)-\(r_h\)) diagrams resemble those characteristic of a Hawking-Page transition. Specifically, a minimum temperature exists below which only thermal AdS is stable; above this temperature, both small and large black hole branches emerge. In contrast, for the standard Reissner-Nordström-AdS black hole, such behavior occurs only in the limit \(Q = 0\).

This phenomenology is governed by two distinct thermodynamic regimes, which we will analyze topologically in the next section.

\subsubsection{Large Charge Regime}
\label{lc}
A standard thermodynamic analysis reveals that for Maxwell charges above \(Q \gtrsim 0.12\) at a coupling constant of \(\lambda = 0.001\), the system demonstrates behavior characteristic of a classical van der Waals fluid \cite{Sadeghi2025}.

For the fixed parameter \(\lambda = 0.001\), numerical evaluation provides the critical parameters detailed in Table~\ref{table1}. This confirms the existence of thermodynamic critical points, defined by the simultaneous conditions \(\partial P/\partial r_h = 0\) and \(\partial^2 P/\partial r_h^2 = 0\), for charges \(Q \gtrsim 0.12\). For charge values below this approximate threshold, these conditions yield no physically admissible real solutions.

\begin{table}[htbp]
    \centering
    \caption{Critical thermodynamic parameters for the van der Waals-like regime at fixed coupling \(\lambda = 0.001\). The critical horizon radius \(r_c\), temperature \(T_c\), and pressure \(P_c\) are shown for selected Maxwell charges \(Q \gtrsim 0.12\).}
    \label{table1}
    \renewcommand{\arraystretch}{1.2}
    \setlength{\tabcolsep}{12pt}
    \begin{tabular}{c c c c}
        \toprule
        \(Q\) & \(r_c\) & \(T_c\) & \(P_c\) \\
        \midrule
        0.20 & 0.3336 & 0.3118 & 0.1729 \\
        0.30 & 0.5115 & 0.2057 & 0.0750 \\
        0.40 & 0.6869 & 0.1537 & 0.0418 \\
        0.50 & 0.8613 & 0.1228 & 0.0266 \\
        0.60 & 1.0353 & 0.1022 & 0.0185 \\
        0.70 & 1.2091 & 0.0876 & 0.0135 \\
        0.80 & 1.3827 & 0.0766 & 0.0103 \\
        \bottomrule
    \end{tabular}
\end{table}

\begin{figure}[H]
    \centering
    \subfloat[Swallowtail structure of the Helmholtz free energy \(F\) versus temperature \(T\) for \(P = 0.01 < P_c\), \(Q = 0.5\), \(\lambda = 0.001\). The crossing of branches indicates a first-order SBH/LBH transition.\label{fig:F_T}]{
        \includegraphics[width=7cm]{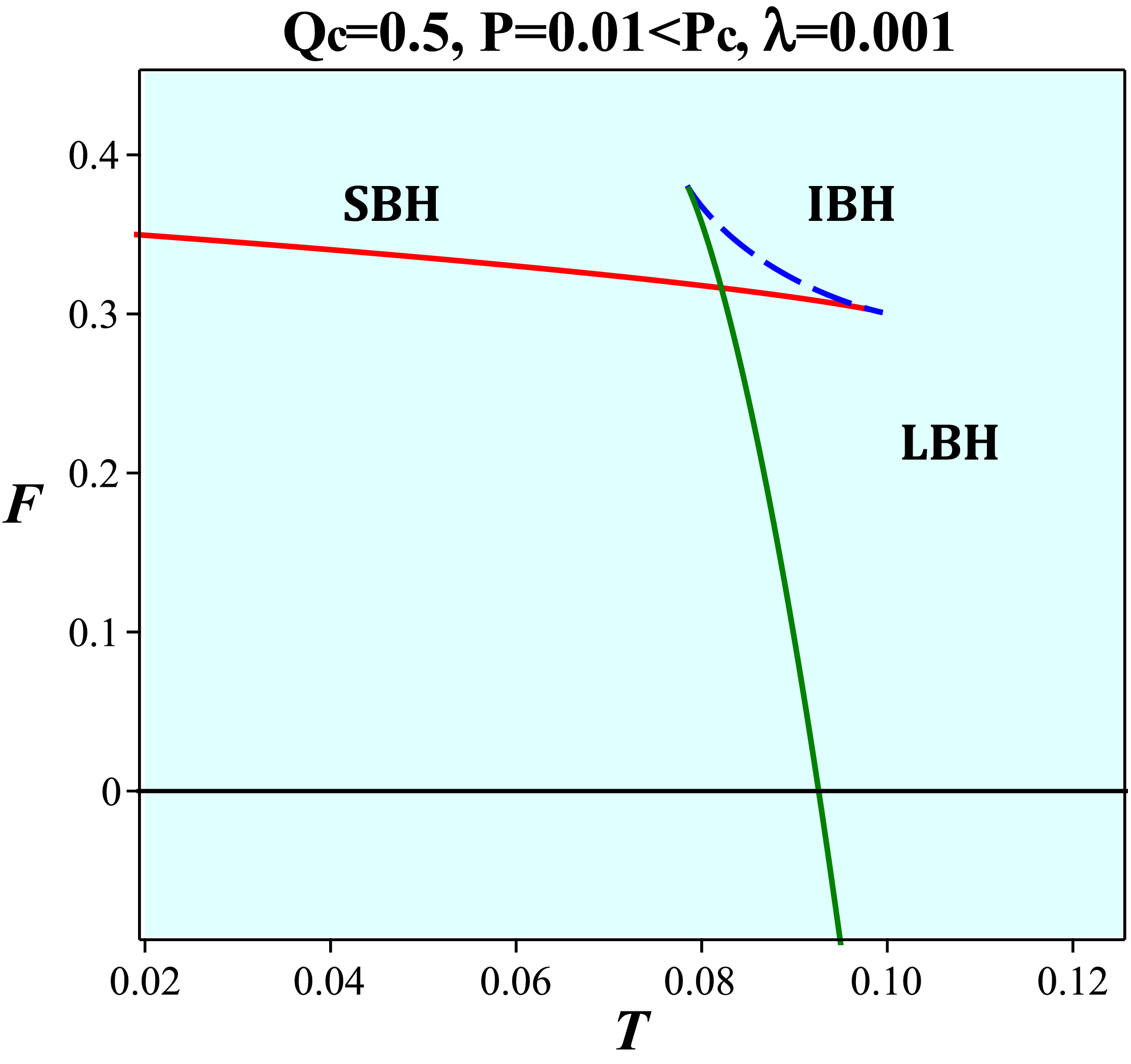}
    }\qquad
    \subfloat[Entropy \(S\) versus temperature \(T\) for the same parameters, showing a discontinuous jump at the transition temperature, which signifies finite latent heat.\label{fig:S_T}]{
        \includegraphics[width=6.5cm]{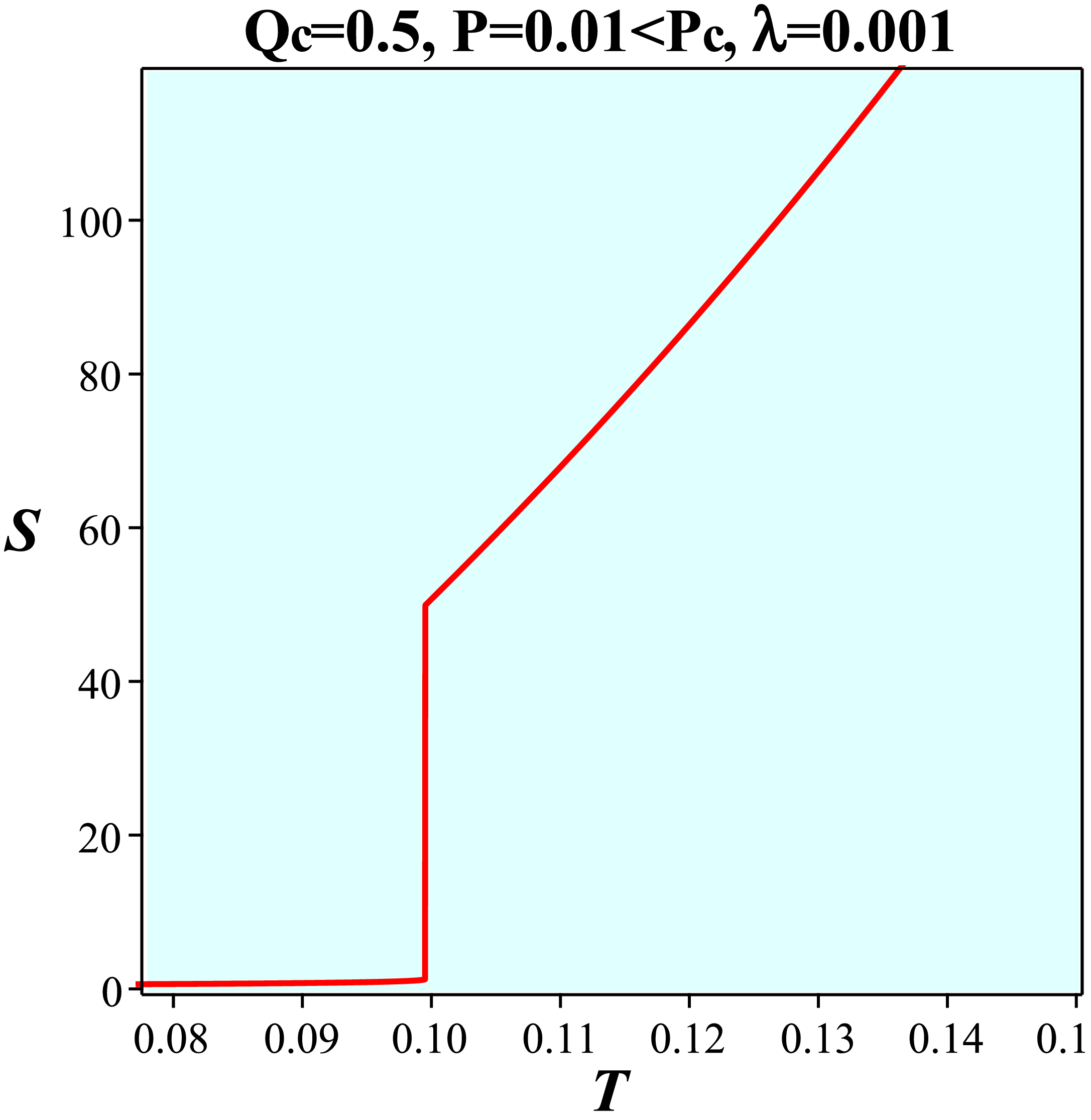}
    }
    \caption{Thermodynamic potentials demonstrating first-order phase transition signatures in the van der Waals regime. The swallowtail structure in \(F(T)\) and the discontinuity in \(S(T)\) occur at subcritical pressure \(P = 0.01\) for \(Q = 0.5\), \(\lambda = 0.001\).}
    \label{fig:combined_plots}
\end{figure}

The system's van der Waals-like character is evidenced by several key thermodynamic features:

\begin{itemize}
    \item \textbf{Isotherms and phase behavior}: The \(P\)–\(r_h\) isotherms exhibit characteristic oscillatory behavior below the critical temperature. In Fig.~\ref{fig:P_rh}, the red curve corresponds to the critical isotherm at \(T_c\) for \(Q=0.5\) (\(T_c=0.1228\)). For comparison, the isotherm at a smaller charge (\(Q=0.1\)) resembles that of a Hawking-Page-type transition. Similarly, in the \(T\)–\(r_h\) diagram (Fig.~\ref{fig:T_rh}), the profile for \(Q=0.1\) displays the characteristic form associated with Hawking-Page transitions.

\begin{figure}[h!]
\centering
\subfloat[Pressure–horizon radius (\(P\)–\(r_h\)) isotherms for selected Maxwell charges. Curves for \(Q = 0.8, 0.5, 0.3\) exhibit van der Waals-like oscillatory behavior, while the \(Q = 0.1\) curve shows Hawking-Page-like character. The red curve marks the critical isotherm for \(Q=0.5\) (\(T_c=0.1228\)).\label{fig:P_rh}]{\includegraphics[width=7cm]{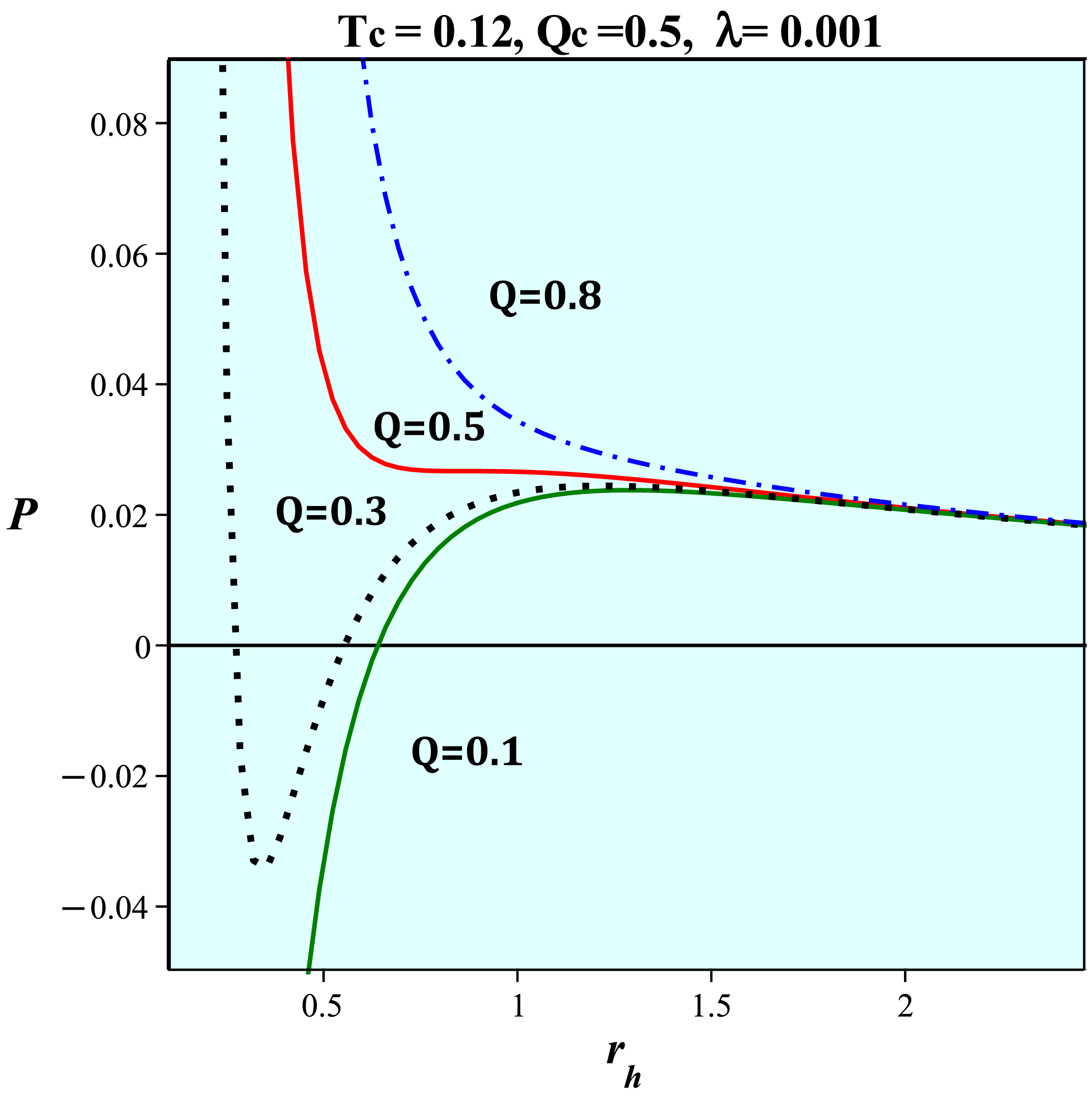}}
\qquad
\subfloat[Temperature–horizon radius (\(T\)–\(r_h\)) relations. Profiles for \(Q = 0.8, 0.5\) show van der Waals-like inflections, while \(Q = 0.1\) exhibits Hawking-Page-like behavior with a single minimum.\label{fig:T_rh}]{\includegraphics[width=7cm]{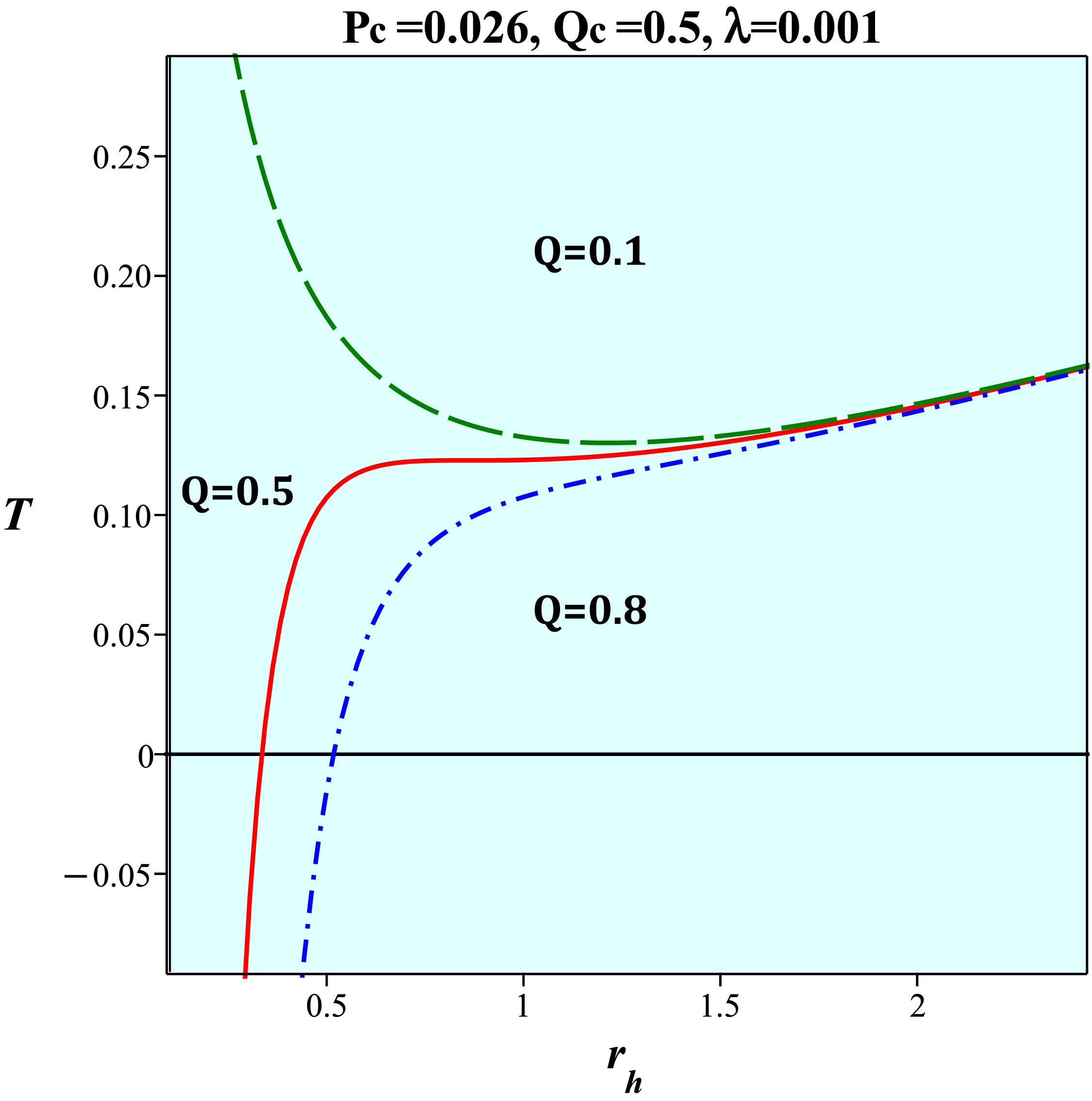}}
\caption{Equation of state and thermal profiles illustrating the crossover from van der Waals to Hawking-Page behavior as the Maxwell charge \(Q\) decreases. The coupling is fixed at \(\lambda = 0.001\).}
\end{figure}

    \item \textbf{Free energy and first-order transitions}: The Helmholtz free energy \(F\) develops a characteristic swallowtail structure (Fig.~\ref{fig:F_T}) at subcritical pressure (\(P = 0.01 < P_c\)), signaling a first-order small-black-hole to large-black-hole (SBH/LBH) phase transition. Here, \(P_c = 0.0266\) is the critical pressure for \(Q=0.5\), \(\lambda=0.001\).
    \item \textbf{Entropy and latent heat}: The corresponding entropy \(S\) displays a discontinuous jump at the transition temperature (Fig.~\ref{fig:S_T}), confirming the presence of latent heat for these parameters.
\end{itemize}

\begin{figure}[H]
\centering
\includegraphics[width=0.4\textwidth]{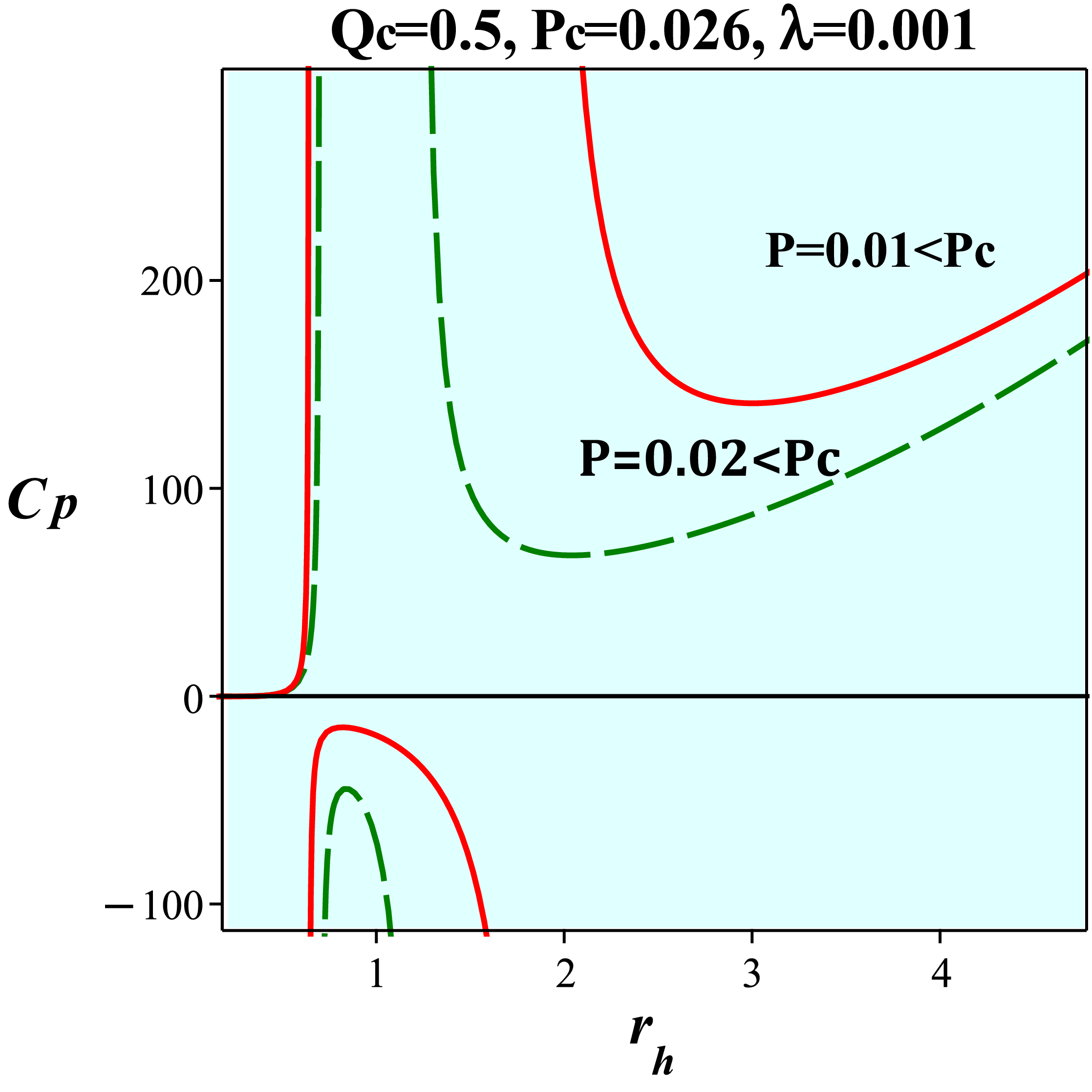}
\caption{Heat capacity at constant pressure, \(C_P\), versus horizon radius \(r_h\) in the van der Waals regime. For subcritical pressure (\(P < P_c\)), the system progresses through two stable phases (positive \(C_P\)) separated by an intermediate unstable phase (negative \(C_P\)).}
\label{fig:C_rh}
\end{figure}

    \item \textbf{Heat capacity and stability}: The heat capacity at constant pressure, \(C_P\), is particularly significant as its sign changes are directly linked to the winding number in the topological approach. For pressures below \(P_c\), \(C_P\) exhibits a sequence of three phases along the \(r_h\) axis: stable (positive), unstable (negative), and stable (positive), as shown in Fig.~\ref{fig:C_rh}.

We will verify this thermodynamic structure through a topological investigation in the following section, which provides a complementary perspective by mapping the phase structure to topological defects in parameter space.

\subsubsection{Small Charge Regime}
\label{sc}
For small Maxwell charges (\(Q \lesssim 0.12\) with \(\lambda = 0.001\)), the system exhibits characteristics consistent with a Hawking-Page transition. In contrast to the van der Waals fluid behavior observed in the large-\(Q\) regime, no physically admissible real roots satisfy the standard critical point conditions \(\partial P/\partial r_h = 0\) and \(\partial^2 P/\partial r_h^2 = 0\).

\begin{figure}[H]
    \centering
    \subfloat[Hawking-Page structure in the Helmholtz free energy \(F\) for \(Q = 0.01\), \(\lambda = 0.001\). The \(F=0\) line (dashed) marks the transition between thermal AdS and the stable large black hole (LBH). The small black hole (SBH) branch has higher free energy and is unstable.\label{fig:F_HP1}]{\includegraphics[width=7cm]{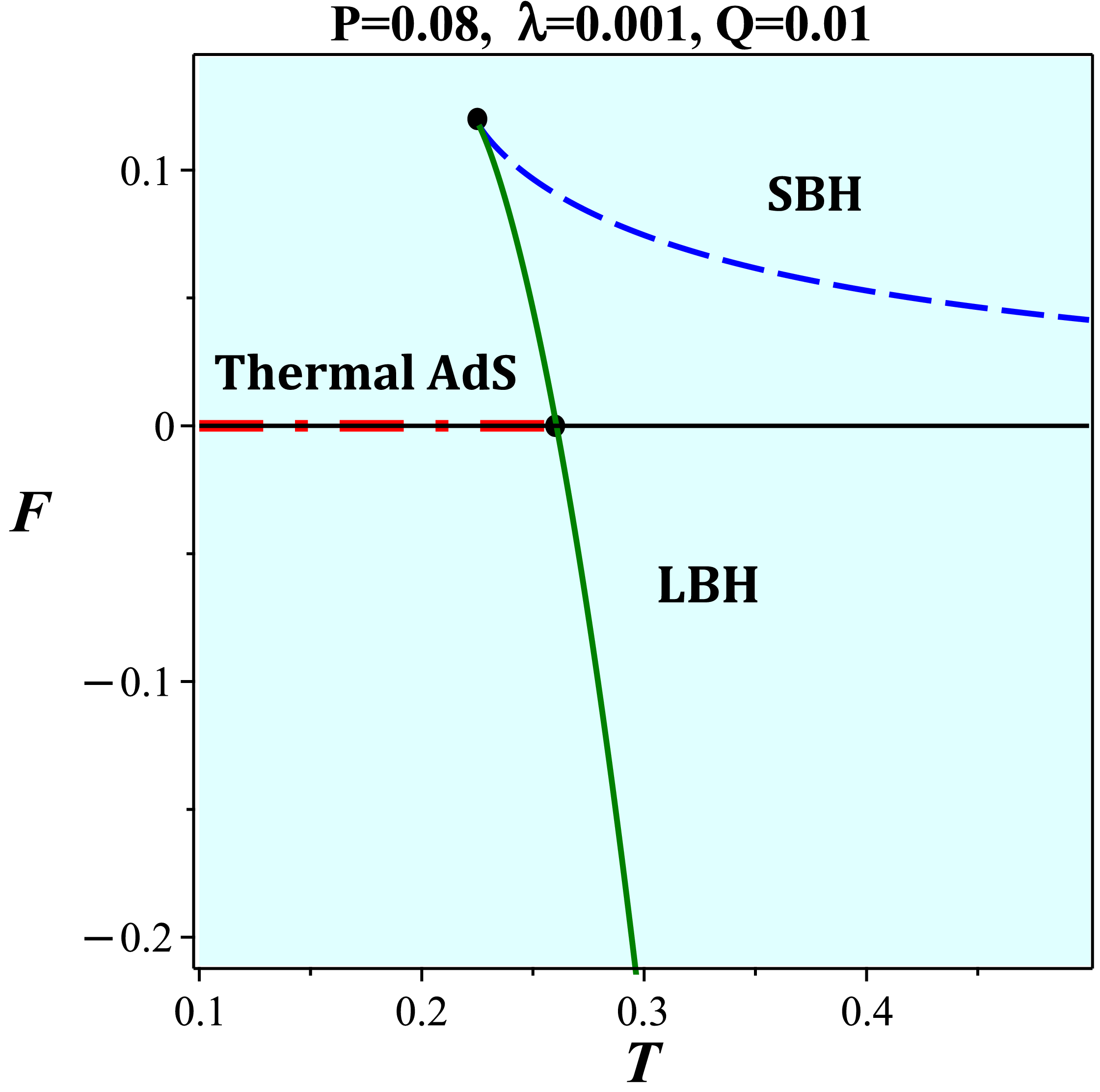}}
    \qquad
    \subfloat[Heat capacity at constant pressure, \(C_P\), for \(Q = 0.01\), \(\lambda = 0.001\), showing the characteristic two-phase structure of the Hawking-Page regime. The SBH branch is unstable (\(C_P < 0\)), while the LBH branch is stable (\(C_P > 0\)).\label{fig:C_HP}]{\includegraphics[width=7cm]{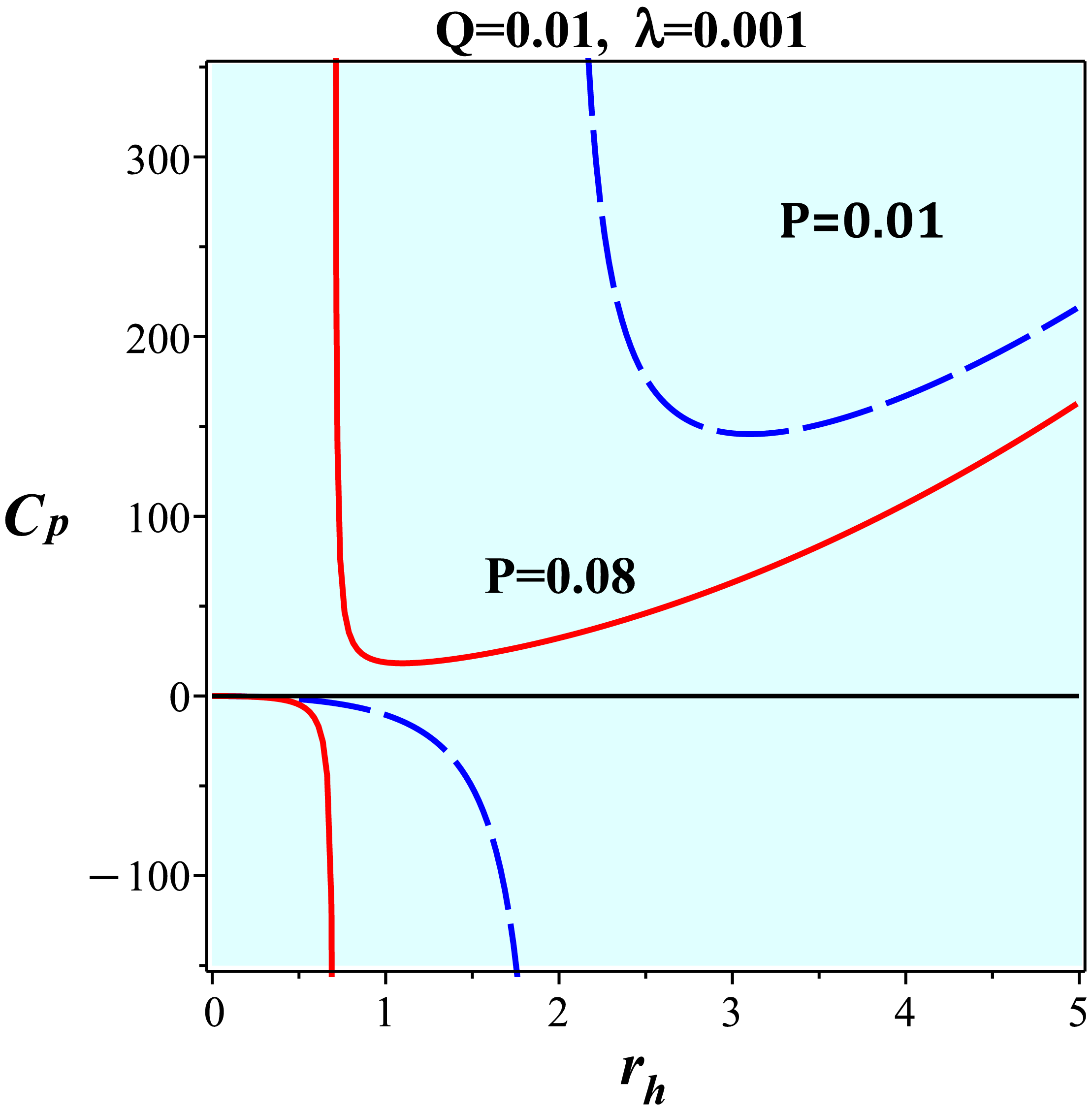}}
    \caption{Thermodynamic potentials in the small-charge (Hawking-Page) regime. Both panels are for \(Q = 0.01\) and \(\lambda = 0.001\), illustrating (a) the global stability condition via free energy and (b) the local stability via heat capacity.}
    \label{fig:combined}
\end{figure}

The salient thermodynamic features of this regime are as follows:

\begin{itemize}
    \item \textbf{Temperature profile and phase existence}: The temperature–horizon radius profiles, \(T(r_h)\), shown in Fig.~\ref{fig:TT_rh}, lack the characteristic van der Waals oscillations. Instead, each profile exhibits a global minimum temperature \(T_{\text{min}}\). For \(T < T_{\text{min}}\), only thermal AdS space is thermodynamically allowed. For \(T > T_{\text{min}}\), two distinct black hole branches coexist: a small black hole (SBH) branch and a large black hole (LBH) branch. While such behavior occurs only at \(Q = 0\) in the standard Reissner–Nordström–AdS black hole, the non-minimal \(\lambda\)-coupling in our model extends this Hawking-Page–like structure to non-zero values of the Maxwell charge.
    
    \begin{figure}[H]
        \centering
        \subfloat[\(Q = 0.01\), \(P=0.08\)\label{fig:t10}]{\includegraphics[width=0.4\textwidth]{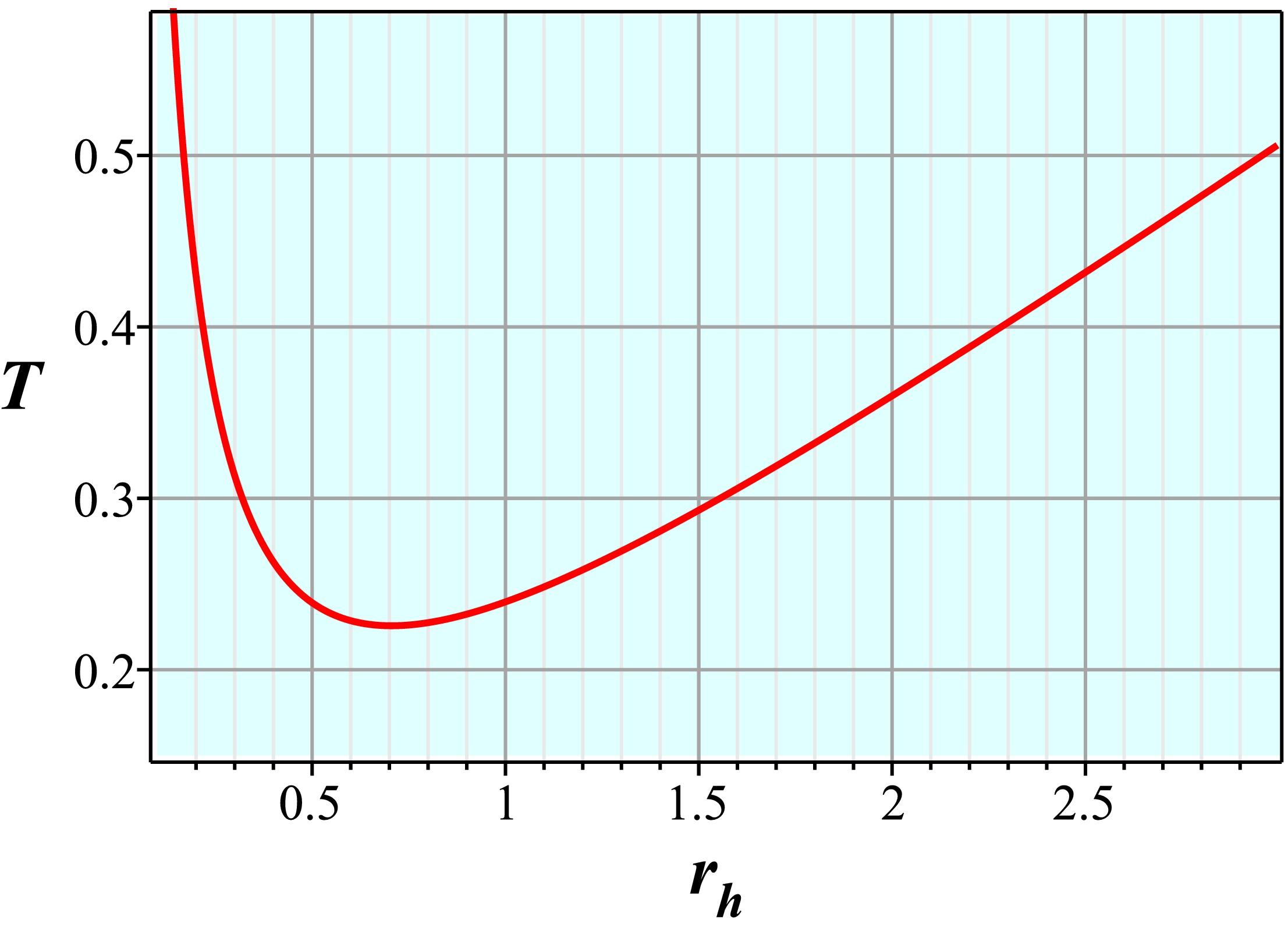}}
        \qquad
        \subfloat[\(Q = 0.05\), \(P=0.08\)\label{fig:t11}]{\includegraphics[width=0.4\textwidth]{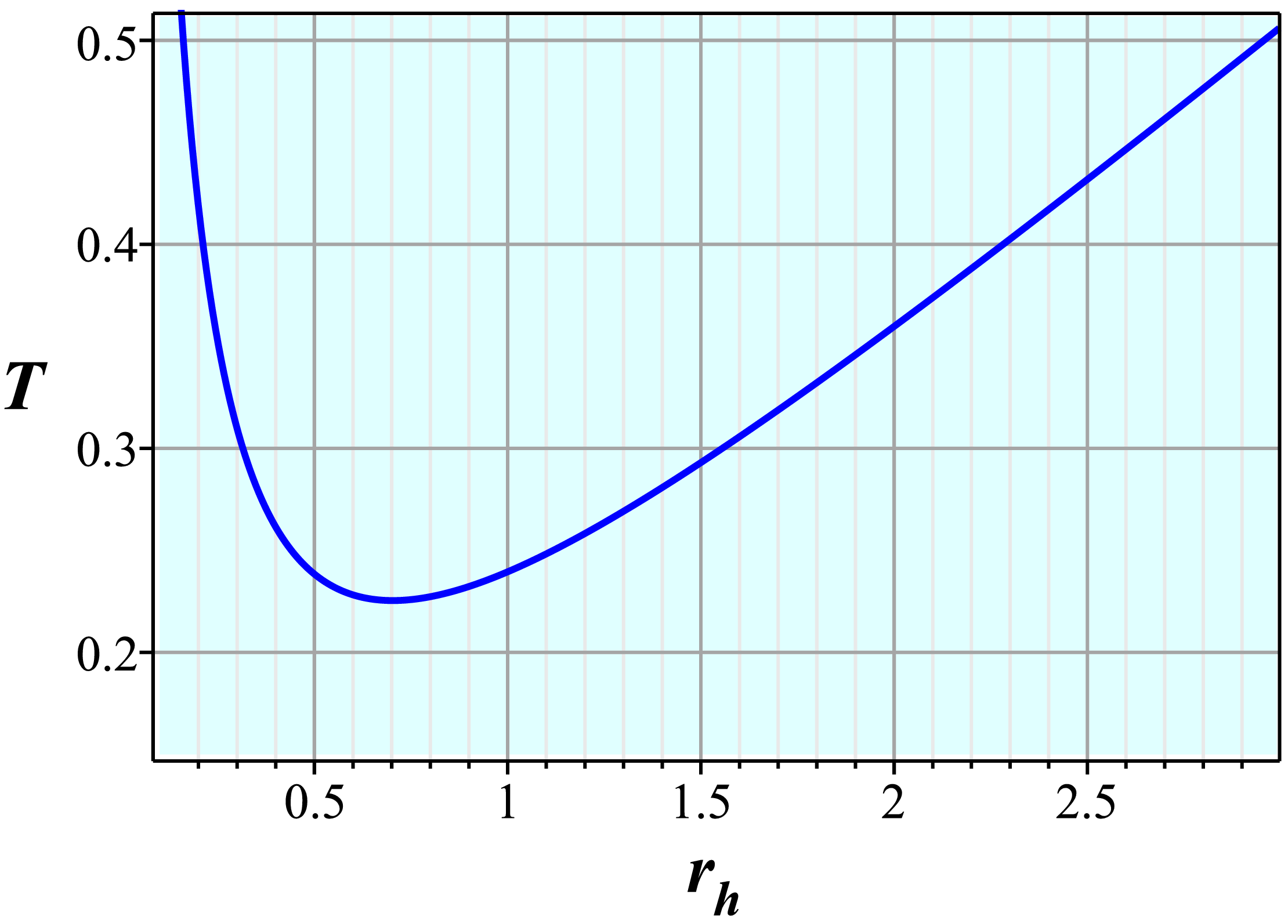}}\\
        \vspace{0.3cm}
        \subfloat[\(Q = 0.1\), \(P=0.01\)\label{fig:t12}]{\includegraphics[width=0.4\textwidth]{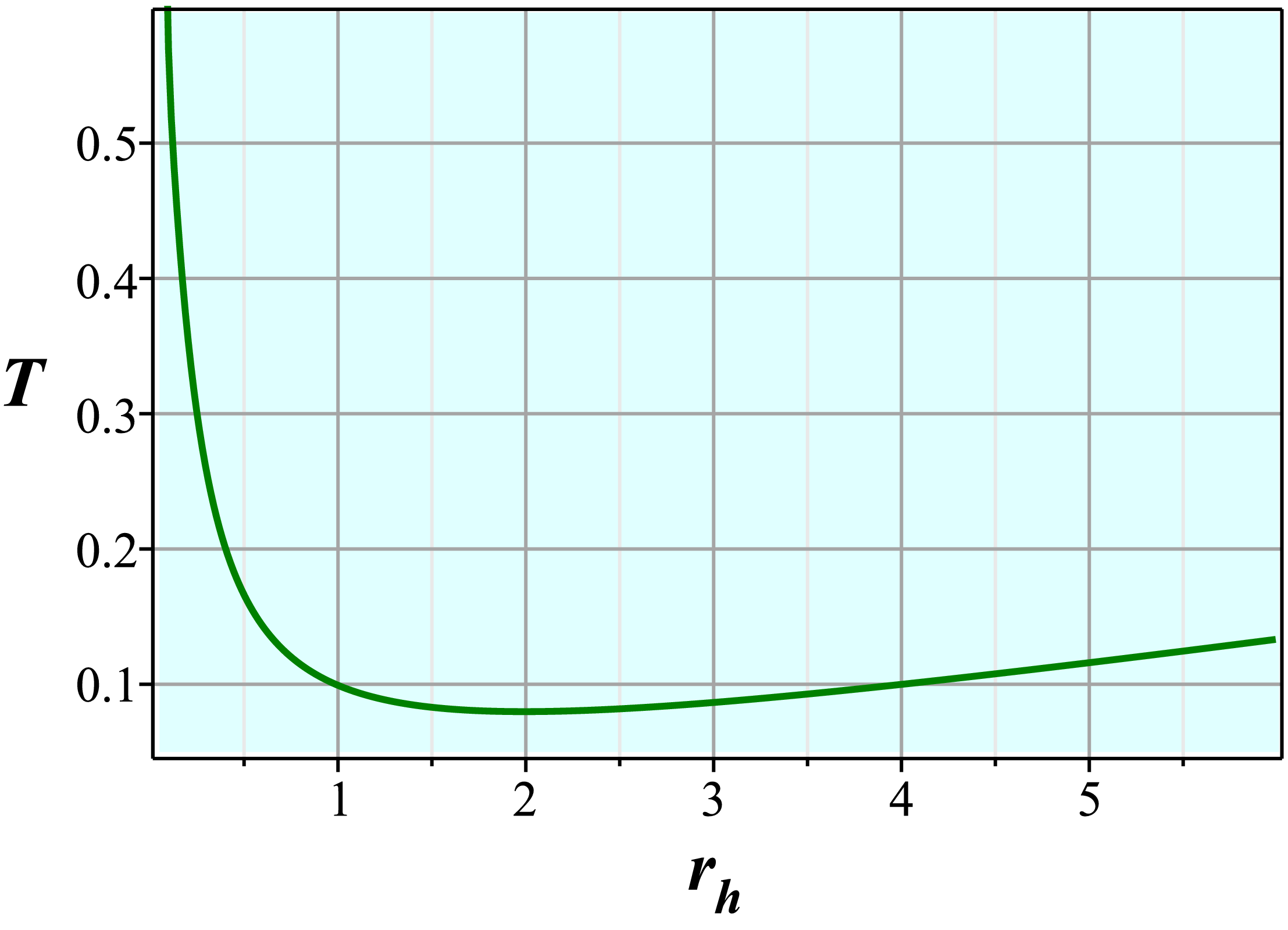}}
        \qquad
        \subfloat[\(Q = 0\), \(P=0.01\)\label{fig:t13}]{\includegraphics[width=0.4\textwidth]{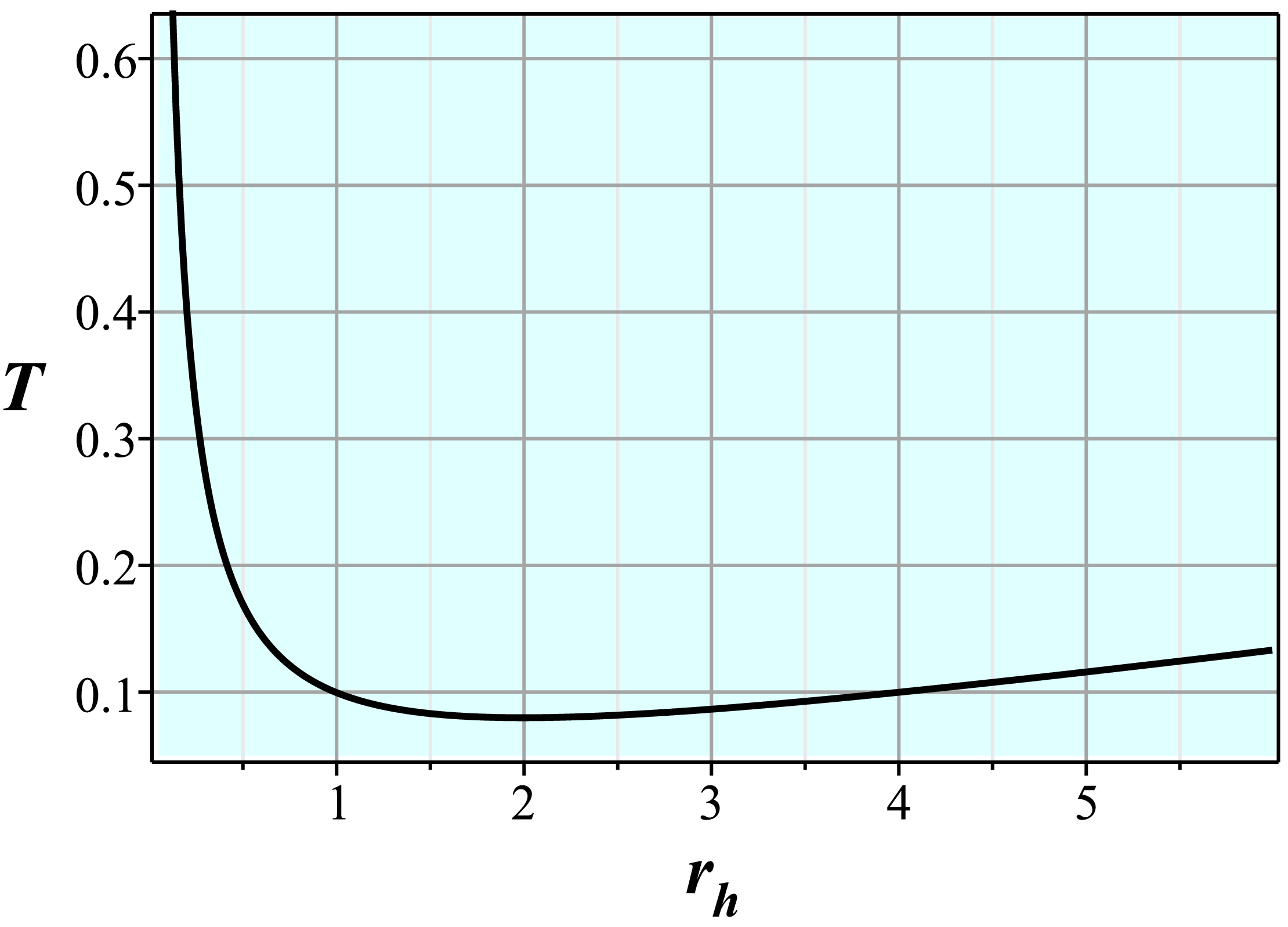}}
        \caption{Temperature versus horizon radius in the small-charge (Hawking-Page) regime. For fixed coupling \(\lambda = 0.001\), profiles are shown for different Maxwell charges \(Q\) and pressures \(P\). Each curve displays a minimum temperature \(T_{\text{min}}\); only thermal AdS exists for \(T < T_{\text{min}}\), while small (SBH) and large (LBH) black hole branches coexist for \(T > T_{\text{min}}\). This structure  demonstrates the persistence of Hawking-Page–like behavior at non-zero charge due to the \(\lambda\)-coupling.}
        \label{fig:TT_rh}
    \end{figure}
        
    \item \textbf{Free energy and global stability}: The Helmholtz free energy \(F(T)\) exhibits the characteristic form of a Hawking-Page transition, as shown in Fig.~\ref{fig:F_HP1}. The point where \(F = 0\) (marked by the horizontal dashed line) defines the phase boundary between stable thermal AdS space and the stable large black hole phase. The small black hole branch, which possesses a higher free energy than both competing states at the same temperature, is globally unstable.
        
    \item \textbf{Heat capacity and local stability}: The heat capacity at constant pressure, \(C_P\), shown in Fig.~\ref{fig:C_HP}, reveals a simple two-phase structure characteristic of Hawking-Page–type systems. The SBH branch has negative \(C_P\) (thermodynamically unstable), while the LBH branch has positive \(C_P\) (thermodynamically stable). This is in contrast to the three-phase structure (\(+,-,+\)) found in the van der Waals regime.
\end{itemize}

The thermodynamic framework established in Sec.~\ref{subsec:thermodynamics} and detailed in this and the preceding subsection provides the necessary foundation for investigating the topological properties of these phase transitions. In summary, a Hawking-Page–type transition dominates for small Maxwell charges, whereas conventional van der Waals fluid behavior is recovered for large \(Q\). In the following section, we will employ a topological approach to independently confirm this dual phase structure and classify its critical points.

\section{Topological Classification of Phase Transitions}
\label{sec3}

Recent advances in black hole thermodynamics have introduced a powerful topological perspective, offering a model-independent framework for classifying thermodynamic phase structures ~\cite{Wei:2022dzw,Wei:2024gfz,Wei2022,Yerra2022}. This approach transcends the details of any specific gravitational theory by focusing on universal, global properties. The core idea is to treat black hole solutions as topological defects within an abstract thermodynamic parameter space \cite{Wei:2022dzw}. This conceptual shift enables a classification based on integer topological invariants, quantities that remain unchanged under continuous deformations of the system's parameters, thereby revealing intrinsic thermodynamic features independent of microscopic details.

\subsection{Foundations of the Topological Approach}
The mathematical foundation of this method originates from topological current theory \cite{Duan1984}. A key element is the construction of a conserved topological current (often referred to as Duan's topological current). For a system described by a two-component vector field $\phi^a(x^\nu)$, this current is defined as
\begin{equation}
j^\mu = \frac{1}{2\pi} \epsilon^{\mu\nu\rho} \epsilon_{ab} \, \partial_\nu n^a \, \partial_\rho n^b,
\end{equation}
where $\partial_\nu \equiv \partial/\partial x^\nu$, the coordinates $x^\nu = (\tau, r_h, \Theta)$ span an extended thermodynamic space, and $n^a = \phi^a / \lVert \phi \rVert$ (with $a=1,2$) is the associated unit vector. Here, $\phi^1 \equiv \phi^{r_h}$ and $\phi^2 \equiv \phi^{\Theta}$. This current is identically conserved, $\partial_\mu j^\mu = 0$. Its structure implies that the charge density $j^0$ is non-vanishing only at the zero points of $\phi^a$, where it takes the form
\begin{equation}
j^0 = \sum_{i=1}^{N} \beta_i \, \eta_i \; \delta^2(\vec{x} - \vec{z}_i).
\end{equation}
The positive integer $\beta_i$ (the Hopf index) counts the number of loops traced by $\phi^a$ in its internal space when $x^\mu$ encircles the zero point $z_i$ of the field i.e $\phi(\vec{x})\vert_{\vec{x}=\vec{z}_i}=0$. The Brouwer degree $\eta_i = \mathrm{sign}\left(J^0(\phi/x)_{z_i}\right) = \pm 1$ indicates the orientation of the vector field at the zero. The integer $N$ is the total number of isolated zero points within the region of interest.

\subsection{Generalized Free Energy and the Vector Field}
Applying this formalism to black hole thermodynamics requires a suitable free energy functional. We build upon the Euclidean quantum gravity approach of Gibbons and Hawking \cite{Gibbons1977}, where the partition function is expressed as a gravitational path integral. York's later refinement \cite{York1986}, placing the black hole in a finite cavity with fixed boundary temperature, stabilizes the canonical ensemble. Within this framework, we introduce a \textit{generalized free energy}
\begin{equation}
\mathcal{F} = \mathcal{H} - \frac{S}{\tau},
\end{equation}
which depends on the black hole horizon radius $r_h$ and an independent temperature parameter $\tau$ (the inverse of the boundary temperature). This free energy is “off-shell” except when $\tau = T^{-1}$, where $T$ is the Hawking temperature of an on-shell black hole solution.

A standard construction for the vector field $\bm{\phi}$ is \cite{Wei2022}
\begin{equation}
\phi = \left( \frac{\partial \mathcal{F}}{\partial r_h},\; -\cot\Theta\,\csc\Theta \right)
      = \left( \frac{\partial \widetilde{\mathcal{F}}}{\partial r_h},\; \frac{\partial \widetilde{\mathcal{F}}}{\partial \Theta}  \right),
\end{equation}
where $\widetilde{\mathcal{F}} \equiv \mathcal{F} + \csc\Theta, 0<r_h<\infty$ and $0<\Theta<\pi$. The angular component is chosen to diverge at $\Theta = 0$ and $\Theta = \pi$, ensuring a consistent, non-vanishing orientation of the vector field on the boundaries of the $(\Theta, r_h)$ plane. The zero points of $\bm{\phi}$, located at $\Theta = \pi/2$ and where $\partial \mathcal{F}/\partial r_h = 0$, correspond precisely to on-shell black hole solutions.

Each zero point is assigned a local topological charge, the winding number $w_i$, which may be positive or negative. Crucially, the sign of $w_i$ encodes the local thermodynamic stability of the corresponding black hole branch: a positive winding number ($w_i = +1$) corresponds to a thermodynamically stable branch (positive heat capacity $C_P > 0$), whereas a negative winding number ($w_i = -1$) indicates a locally unstable branch ($C_P < 0$).

The sum of the winding numbers for all black hole branches at a given temperature yields a global topological invariant, the \textit{total topological number}:
\begin{equation}
W = \int_{\Sigma} j^0 \, d^2 x = \sum_{i=1}^{N} \beta_i \eta_i = \sum_{i=1}^{N} w_i.
\end{equation}
The winding number for a zero point enclosed by a contour $C_i$ is defined mathematically as
\begin{equation}\label{winint}
w_i = \frac{1}{2\pi} \oint_{C_i} d\Omega, 
\end{equation}
where $\Omega$ is the phase angle of the vector $\bm{\phi}$ in the  $\phi$-plane\cite{Liu:2022aqt}.

\subsection{Topological Classes and Physical Interpretation}
The integer $W$ classifies the global thermodynamic structure of black hole systems and provides deep insight into their permissible phase behavior. In particular:
\begin{itemize}
    \item A topological class with $W = -1$ indicates the presence of only unstable thermodynamic branches, ruling out any equilibrium phase transitions.
    \item A class with $W = 0$ corresponds to systems with balanced stable and unstable branches. This is characteristic of Hawking–Page–type transitions, as seen in Schwarzschild–AdS black holes.
    \item A class with $W = +1$ signifies a dominance of stable branches, typically associated with systems exhibiting van der Waals–type liquid–gas phase transitions, such as charged AdS black holes.
\end{itemize}
Remarkably, $W$ depends only on the asymptotic behavior of the inverse temperature $\beta(r_h) \equiv 1/T(r_h)$ in the limits of small and large horizon radii, making it a robust and universal feature of the solution. Further details of the classification are discussed in Ref.~\cite{Wei:2024gfz}.

\subsection{Asymptotic Analysis and Boundary Behavior}
The classification hinges on the asymptotic behavior of the components of the vector field  along the boundaries of the $(r_h, \Theta)$ plane~\cite{Wei:2024gfz}. We consider a rectangular contour $C = I_1 \cup I_2 \cup I_3 \cup I_4$ defined by
\begin{align}
I_1 &= \{r_h = \infty,\; \Theta \in (0, \pi)\}, \nonumber \\
I_2 &= \{r_h \in (\infty, r_m),\; \Theta = \pi\}, \nonumber \\
I_3 &= \{r_h = r_m,\; \Theta \in (\pi, 0)\}, \nonumber \\
I_4 &= \{r_h \in (r_m, \infty),\; \Theta = 0\}. \nonumber
\end{align}
Here, $r_m$ denotes the minimal possible horizon radius for a physical black hole solution, defining the lower bound of the physical domain $(r_m, \infty)$.

The behavior of $\phi^{r_h}$ on $I_1$ and $I_3$ follows from the definition of the generalized free energy. Starting from $\phi^{r_h} = \partial \widetilde{\mathcal{F}}/\partial r_h = \partial \mathcal{F}/\partial r_h$ and using $\mathcal{F} = \mathcal{H} - S/\tau$, we have
\begin{equation}\label{phid}
\phi^{r_h} = \frac{\partial \mathcal{H}}{\partial S} \frac{\partial S}{\partial r_h} - \frac{1}{\tau} \frac{\partial S}{\partial r_h}.
\end{equation}
Identifying $\partial \mathcal{H}/\partial S = T = 1/\beta$, where $\beta$ is the inverse Hawking temperature of an on-shell black hole with horizon radius $r_h$, Eq.~\eqref{phid} simplifies to
\begin{equation}
\phi^{r_h} = \frac{\partial S}{\partial r_h} \left( \frac{1}{\beta} - \frac{1}{\tau} \right).
\end{equation}
Since $\tau$ is the fixed cavity temperature and $\partial S/\partial r_h > 0$ for black holes, the sign of $\phi^{r_h}$ is determined solely by the difference $(1/\beta - 1/\tau)$.

For the Reissner–Nordström–AdS (RN-AdS) black hole, $r_m$ is the extremal horizon radius (positive, determined by $Q$ and the AdS length $l$). At extremality, $T(r_m)=0$, so $\beta(r_m) \to \infty$. For the Schwarzschild–AdS (SAdS) black hole, there is no extremal limit from charge; one may take $r_m \to 0$ thermodynamically. In this small-horizon limit, $T(r_h)$ diverges, giving $\beta(r_m) \to 0$.

This leads to two distinct asymptotic patterns:
\begin{itemize}
    \item \textbf{Case III (RN-AdS type):} $\beta(r_m) = \infty$, $\beta(\infty) = 0$. Consequently, as $r_h \to r_m$, $\phi^{r_h} \to -$ (points left, $\leftarrow$); as $r_h \to \infty$, $\phi^{r_h} \to +$ (points right, $\rightarrow$).
    \item \textbf{Case IV (SAdS type):} $\beta(r_m) = 0$, $\beta(\infty) = 0$. Here, $\phi^{r_h} \to +$ (points right, $\rightarrow$) both as $r_h \to r_m$ and as $r_h \to \infty$.
\end{itemize}

The complete boundary behavior for these two paradigmatic cases is summarized in Table~\ref{boundary}. The arrows indicate the direction of the $\bm{\phi}$-field vector along each contour segment.

\begin{table}[h]
\centering
\caption{Boundary behavior of $\bm{\phi}$ and topological number $W$ for RN-AdS and Schwarzschild-AdS black holes.}
\label{boundary}
\begin{tabular}{lccccc}
\toprule
\textbf{Black Hole} & \textbf{$I_1$} & \textbf{$I_2$} & \textbf{$I_3$} & \textbf{$I_4$} & \textbf{$W$} \\
& $(r_h \to \infty)$ & $(\Theta=\pi)$ & $(r_h \to r_m)$ & $(\Theta=0)$ & \\
\midrule
RN-AdS (Case III) & $\rightarrow$ & $\uparrow$ & $\leftarrow$ & $\downarrow$ & $+1$ \\
Schwarzschild-AdS (Case IV) & $\rightarrow$ & $\uparrow$ & $\rightarrow$ & $\downarrow$ & $0$ \\
\bottomrule
\end{tabular}
\end{table}

For RN-AdS, the typical configuration involves three zero points (branches): a stable small black hole ($w=+1$), an intermediate unstable black hole ($w=-1$), and a stable large black hole ($w=+1$), summing to $W = +1$. For SAdS, there are two zero points: an unstable small black hole ($w=-1$) and a stable large black hole ($w=+1$), giving $W = 0$.

This topological classification partitions black hole systems into universality classes characterized by their $W$ number, which fundamentally constrains their possible phase structures and stability properties. It provides a powerful, universal framework for understanding black hole thermodynamics across different theories of gravity.

In the following subsections, we apply this topological approach to our specific model of a non-minimally coupled black hole. We will show that our system belongs to Case III or Case IV depending on the Maxwell charge $Q$, thereby revealing its intrinsic thermodynamic character through the lens of topological classification and connecting it directly to the phase behavior identified in Sec.~\ref{lc} and Sec.~\ref{sc}.
 
\subsection{Topological Study for Large $Q$ Regime}
\label{tlc}

A standard thermodynamic analysis, detailed in the previous section, reveals that for large Maxwell charges (\(Q \gtrsim 0.12\)), the system exhibits behavior characteristic of a van der Waals fluid. We therefore expect three zero points in the \(r_h\)--\(\Theta\) plane, corresponding to three on-shell black hole solutions. This expectation is evident from the heat capacity diagrams, which indicate two stable phases (positive heat capacity) separated by an unstable phase (negative heat capacity). In the topological framework, this configuration corresponds to two defects with positive winding numbers and one with a negative winding number. We now discuss this scenario in detail.

Within the framework of topological black hole thermodynamics, the number and nature of zero points correspond to on-shell black hole solutions, with their winding numbers encoding local thermodynamic stability. The specific pattern of three zero points with winding numbers \(+1\), \(-1\), and \(+1\) identifies our system as belonging to \textit{Case III} in the asymptotic classification, where the global topological number is \(W = \sum w_i = +1\). This classification originates from the asymptotic behavior of the inverse temperature \(\beta(r_h) = 1/T(r_h)\): as \(r_h \to r_m\) (the extremal limit), \(\beta \to \infty\), and as \(r_h \to \infty\), \(\beta \to 0\). The inverse temperature is obtained from the Hawking temperature relation \eqref{temperature}. Consequently, the innermost (small) and outermost (large) black hole branches are thermodynamically stable (\(w = +1\)), while an intermediate unstable branch (\(w = -1\)) emerges between them. This structure is characteristic of van der Waals-like phase transitions. To confirm this assertion, we investigate and illustrate the topological behavior of the system in this regime.

The off-shell free energy for our system is given by
\begin{equation}\label{eq:F_offshell_largeQ}
F = \mathcal{H} - \frac{S}{\tau} =\frac{r_h}{2}+\frac{4\pi P r_h^3}{3} + \frac{Q^2}{4r_h}+ \left( -\frac{8Q^2\pi P}{r_h} + \frac{Q^4}{5 r_h^5} - \frac{Q^2}{r_h^3} \right)\lambda - \frac{1}{\tau}\left( \pi r_h^2 - \frac{4\pi Q^2\lambda}{r_h^2} \right).
\end{equation}
From this, we obtain the components of the vector field \(\bm{\phi}\):
\begin{align}
\phi^{r_h} &= \frac{\partial F}{\partial r_h} =\frac{1}{2}+4\pi P r_h^2 - \frac{Q^2}{4r_h^2} + \left( \frac{8\pi P Q^2}{r_h^2} - \frac{Q^4}{r_h^6} + \frac{3Q^2}{r_h^4} \right)\lambda - \frac{1}{\tau}\left( 2\pi r_h + \frac{8\pi \lambda Q^2}{r_h^3} \right), \label{eq:phi_rh_largeQ} \\
\phi^{\Theta} &= -\cot\Theta \csc\Theta. \label{eq:phi_Theta_largeQ}
\end{align}
The component \(\phi^{\Theta}\) vanishes at \(\Theta = \pi/2\) and diverges at \(\Theta = 0\) and \(\Theta = \pi\), ensuring a consistent outward orientation of the vector field at the boundaries \cite{Wei:2024gfz}. The boundary behavior is essential for the topological classification, as it fixes the direction of \(\bm{\phi}\) along segments \(I_2\) and \(I_4\) of the parameter space contour. 

Using the information from Table~\ref{table1}, which shows parameter values where van der Waals behavior dominates, we identify the different zero points and calculate their winding numbers. We then illustrate the unit vector field \(n^a = \phi^a/|\phi|\) on a portion of the \(r_h\)--\(\Theta\) plane.

For the Maxwell charge \(Q = 0.5\) and \(\lambda = 0.001\), the critical pressure and temperature are \(P_c \approx 0.026\) and \(T_c \approx 0.12\), respectively. When the free energy becomes on-shell, we have \(\tau_c = 1/T_c \approx 8.33\). For \(\tau > \tau_c\) and pressures below \(P_c\), we observe three defects corresponding to on-shell black hole solutions.

The heat capacity diagrams (Fig.~\ref{fig:C_rh}) suggest that the winding numbers should follow the pattern positive/negative/positive. For the parameter values \(\tau = 11\), \(P = 0.01\), \(Q = 0.5\), and \(\lambda = 0.001\), the zeros occur at \(r_{h1} = 0.500\), \(r_{h2} = 0.959\), and \(r_{h3} = 3.381\). By considering counterclockwise contours around each zero and using integral (\ref{winint}), we find that the winding numbers associated with each zero, ordered from smallest to largest horizon radius, are \(w = (1,\; -1,\; 1)\). To compute these, we can choose parametric equations describing a contour \(C_i\) around a zero point \(z_i\) in the \(r_h\)--\(\Theta\) plane as
\begin{equation}\label{cparm}
r_h = a \cos t + Z_i, \qquad \Theta = b \sin t + \frac{\pi}{2},
\end{equation}
where \(a\) and \(b\) are contour parameters, and evaluate the integral \eqref{winint}.

Equivalently, for isolated non-degenerate zeros, the winding number equals the Brouwer degree \(\eta_i = \text{sign}(J^0(\phi/x))\), where \(J^0\) is the Jacobian determinant of the map \(\bm{\phi}\) \cite{Duan1984,Wei2022}. The sign directly correlates with thermodynamic stability: \(w = +1\) corresponds to positive heat capacity (stable branch), while \(w = -1\) indicates negative heat capacity (unstable branch).

\begin{figure}[H]
\centering
\subfloat[Vector field topology]{\includegraphics[width=7cm]{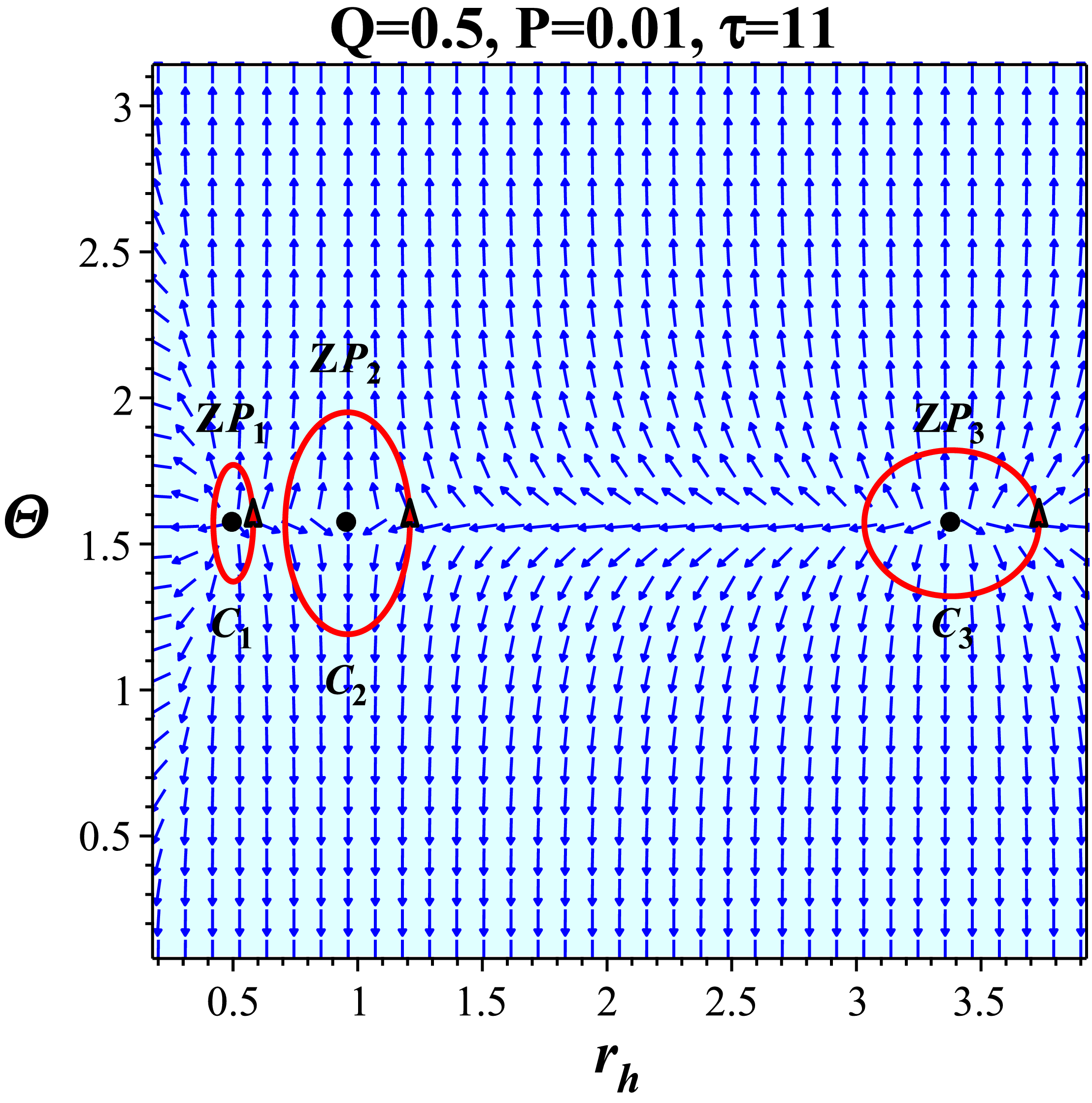}\label{fig:w1}}\qquad
\subfloat[\(r_h\)--\(\tau\) diagram]{\includegraphics[width=7cm]{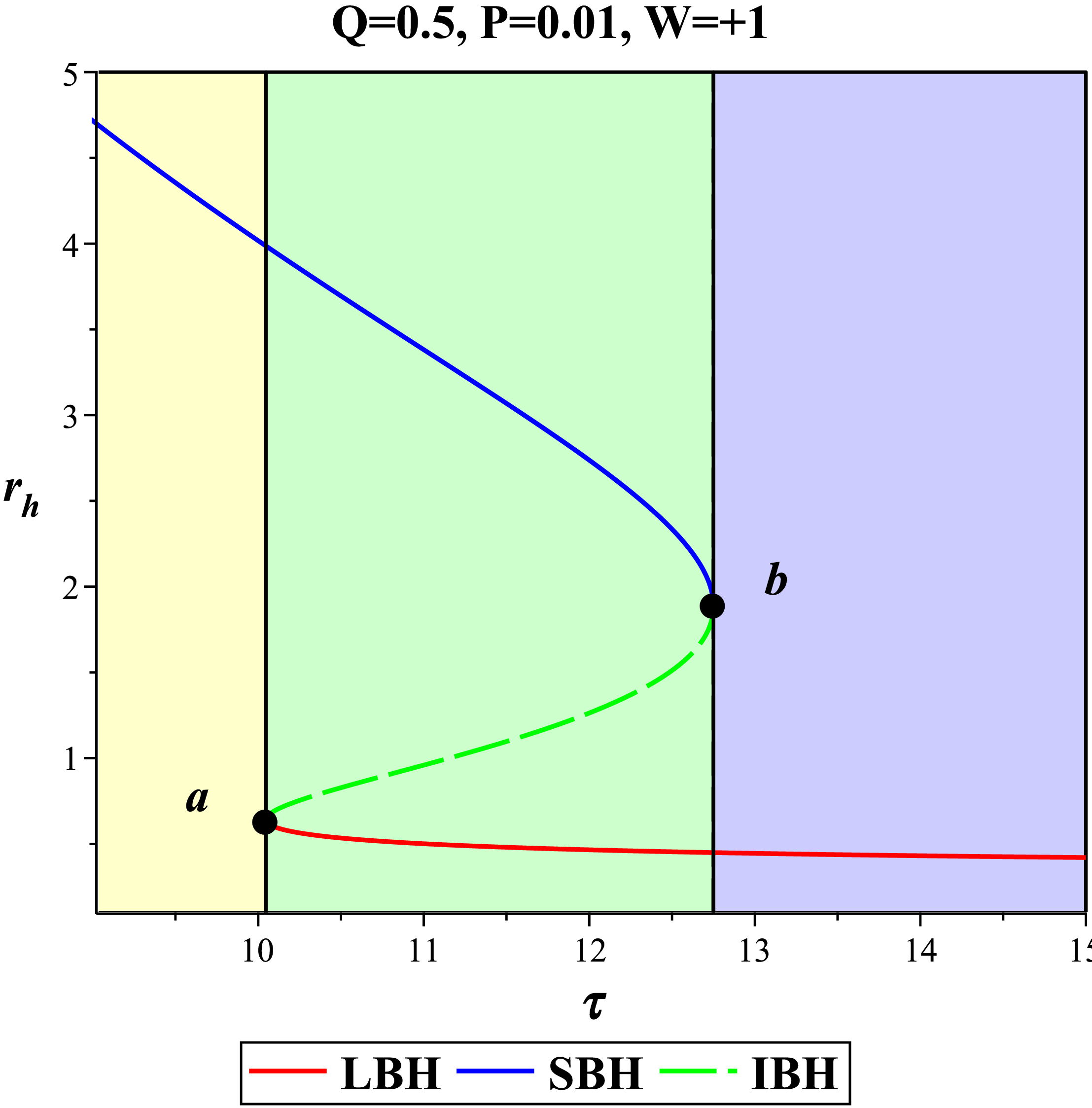}\label{fig:turn1}}
\caption{\textbf{(a)} Vector field topology showing zero points (black dots) at \((r_h,\Theta) = (0.500,\pi/2)\), \((0.959,\pi/2)\), and \((3.381,\pi/2)\) for ZP1, ZP2, and ZP3, respectively. Red contours \(C_i\) enclose these points for \(Q=0.5, P=0.01, \tau=11\). The behavior of the unit vector field at the boundaries shows the system lies within Case III. \textbf{(b)} \(r_h\)--\(\tau\) diagram showing generation and annihilation points in the large-\(Q\) regime. Points \(a\) and \(b\) mark the generation and annihilation of black holes at \(\tau=10.04\) and \(\tau=12.75\), respectively. For each colored rectangular region, \(W=1\).}
\label{fig:wt}
\end{figure}

Figure~\ref{fig:wt}(a) displays the zero points (ZPs) marked with black dots at \((r_h, \Theta) = (0.500, \pi/2)\), \((0.959, \pi/2)\), and \((3.381, \pi/2)\) for ZP1, ZP2, and ZP3, respectively. The red contours \(C_i\) are closed loops enclosing these points for \(\tau = 11\). The boundary behavior of the vector field confirms that the system belongs to Case III of the topological classification mentioned earlier.

We now analyze the \(r_h\)--\(\tau\) diagrams, which reveal the generation and annihilation points of the system. Solving \(\phi^{r_h}=0\) for \(\tau\) yields
\begin{equation}\label{tau5}
\tau = \frac{8\pi \, (r_h^4 + 4Q^2\lambda) \, r_h^3}{16\pi P r_h^8 + 32\pi P Q^2\lambda r_h^4 - Q^2 r_h^4 + 2r_h^6 - 4Q^4\lambda + 12Q^2\lambda r_h^2}.
\end{equation}
Two turning points are observed in the large-\(Q\) regime (see Fig.~\ref{fig:wt}(b)); from bottom to top, these correspond to the generation and annihilation of black hole branches. This curve is plotted using \(\phi^{r_h}=0\) for parameters \(Q=0.5\) and \(P=0.01\).

For \(\tau > 12.75\), only a small black hole exists. For \(\tau < 10.04\), only a large black hole exists. In the interval \(10.04 < \tau < 12.75\), three states—small (SBH), intermediate (IBH), and large (LBH) black holes coexist. However, in each case the total topological number remains invariant (\(W=1\)).

\begin{figure}[H]
\centering
\subfloat[ \(Q=0.5\), \(P=0.007\), \(\tau=15\)]{\includegraphics[width=7cm]{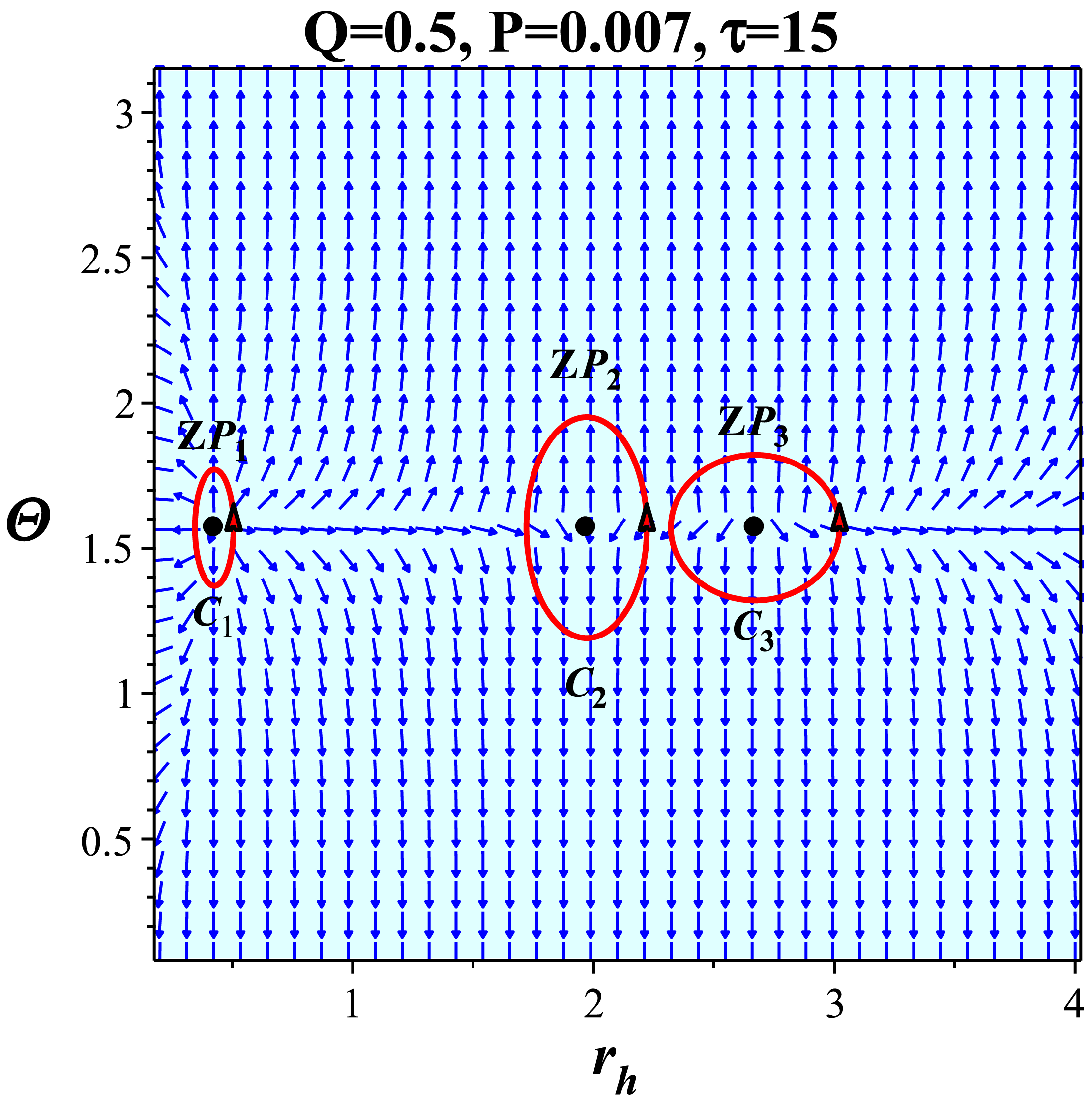}\label{fig:w2}}\qquad
\subfloat[\(Q=0.7\), \(P=0.01\), \(\tau=13\)]{\includegraphics[width=7cm]{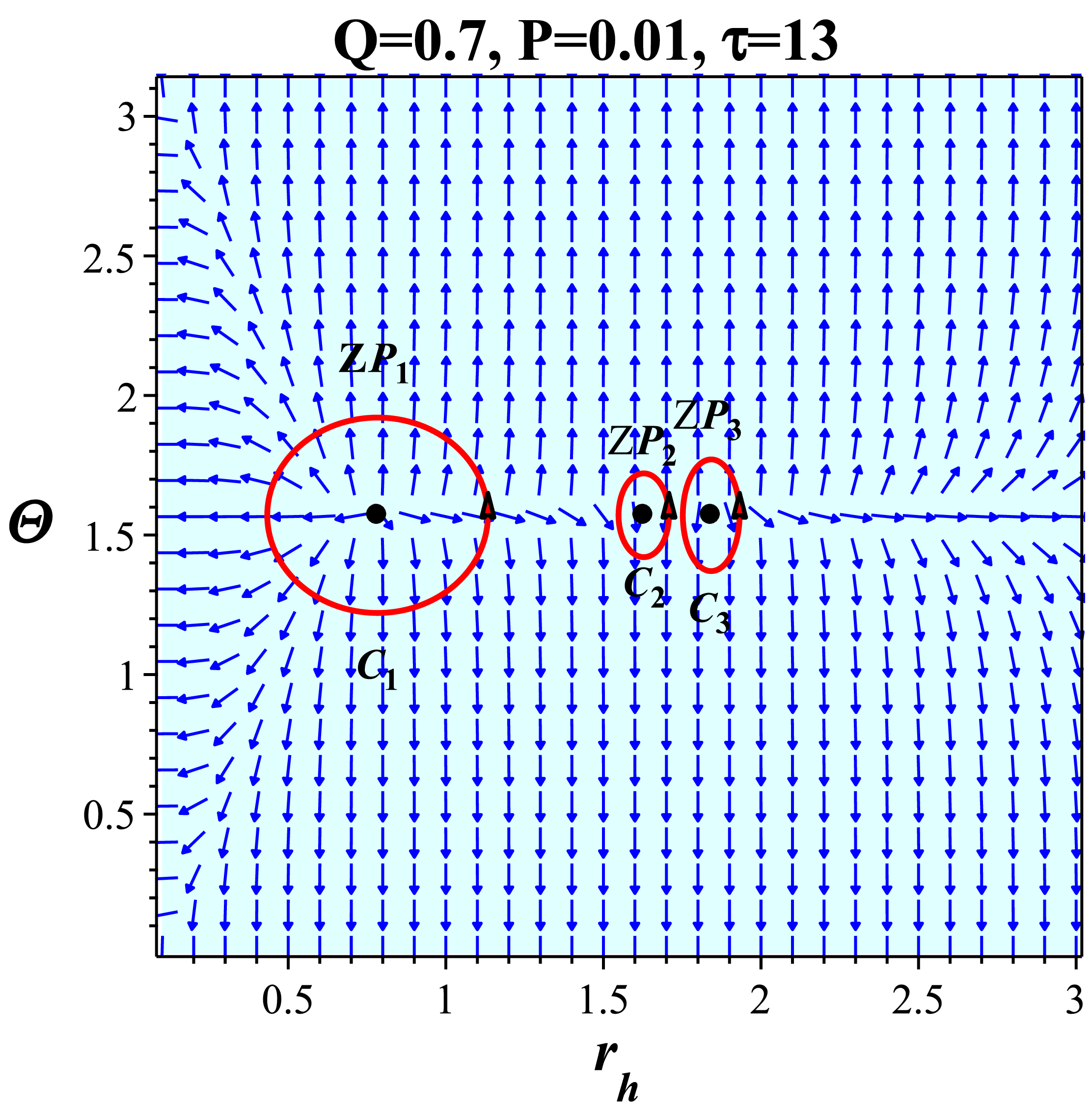}\label{fig:w4}}
\caption{Unit vector field diagrams: \textbf{(a)} Zero points at \((r_h,\Theta) = (0.42,\pi/2)\), \((1.97,\pi/2)\), and \((2.67,\pi/2)\). \textbf{(b)} Zero points at \((r_h,\Theta) = (0.78,\pi/2)\), \((1.62,\pi/2)\), and \((1.84,\pi/2)\). Red contours \(C_i\) enclose these points. The boundary behavior of the vector field places the system within Case III of the topological classification.}
\label{fig:ww2}
\end{figure}

For further verification, we varied several parameters and observed similar behavior for large \(Q\) values. In both panels of Figure~\ref{fig:ww2}, at the boundaries \(I_1\) and \(I_3\) of the \(r_h\)--\(\Theta\) plane, the vector field component \(\phi^{r_h}\) points rightward and leftward, respectively, confirming that our model for large \(Q\) lies in the Case III classification. The left and right panels of Figure~\ref{fig:ww2} show zero points marked with black dots surrounded by red contours \(C_i\) for \(Q=0.5, P=0.007, \tau=15\) and \(Q=0.7, P=0.01, \tau=13\), respectively.

\begin{figure}[H]
\centering
\subfloat[Case: $Q=0.5$, $P=0.01$, $\tau=11$]{\includegraphics[width=7cm]{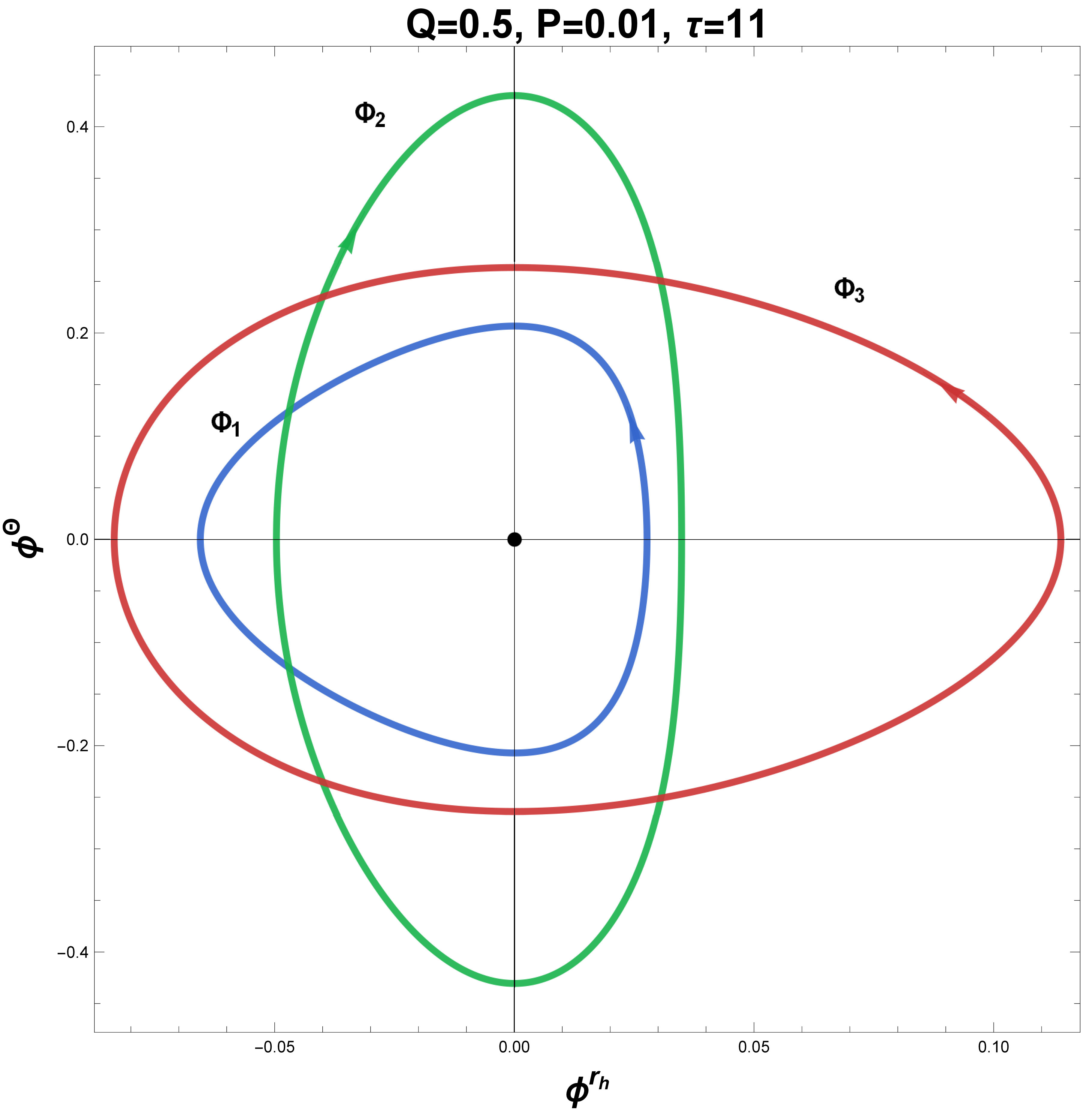}\label{fig:con11}}\qquad
\subfloat[Case: $Q=0.5$, $P=0.007$, $\tau=15$]{\includegraphics[width=7cm]{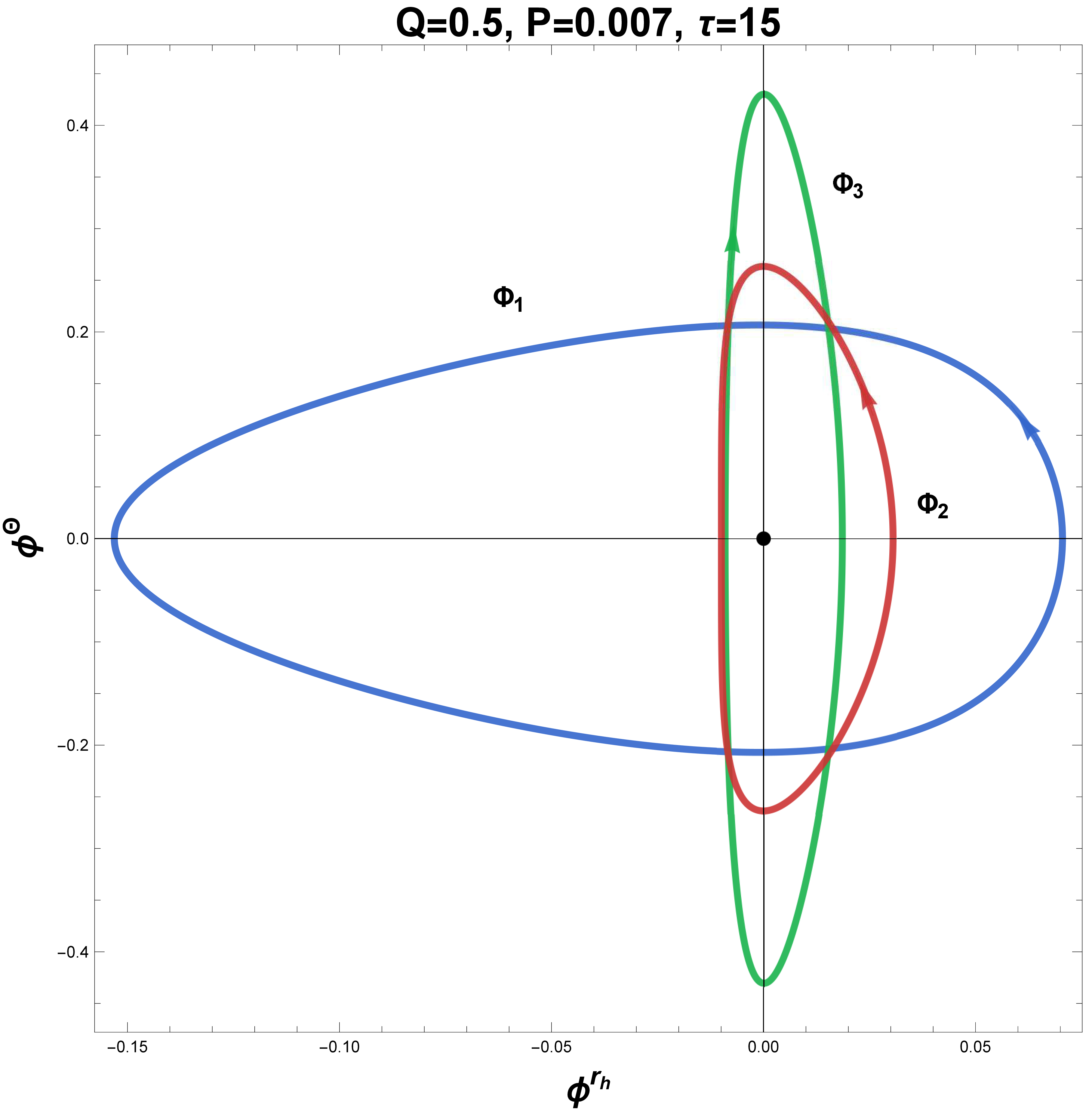}\label{fig:con15}}
\caption{Mapping of contours from the \(r_h\)--\(\Theta\) plane to the \(\phi^{r_h}\)--\(\phi^{\Theta}\) plane, illustrating the winding behavior of $\phi$-field around zero points. The left panel corresponds to the contours in Fig.~\ref{fig:wt}(a); the right panel corresponds to those in the left panel of Fig.~\ref{fig:ww2}.}
\label{fig:con1115}
\end{figure}

Mapping contours from the \(r_h\)--\(\Theta\) plane to the \(\phi^{r_h}\)--\(\phi^{\Theta}\) plane allows us to verify the winding numbers by examining the behavior around the zero point of the field. The direction associated with each contour reveals the topological winding behavior linked to black hole states. Each closed contour \(\Phi_i\) in Fig.~\ref{fig:con1115} shows the change in the components of the \(\bm{\phi}\)-field along the image of a contour encircling a zero point (representing a black hole state) in the \(r_h\)--\(\Theta\) plane. According to the topological dictionary, contours that map to \textit{counterclockwise} loops in the \(\bm{\phi}\)-plane indicate a winding number \(w = +1\) (stable branches), while \textit{clockwise} loops indicate \(w = -1\) (unstable branches). This mapping provides independent geometric verification of our winding number assignments.

The direction of the contours in the \(\bm{\phi}\)-plane can be understood as follows. Consider consecutive points \(p_1, p_2, \ldots, p_n\) along a contour \(C_i\) in the \(r_h\)--\(\Theta\) plane, traversed conventionally counterclockwise. Mapping these points via the \(\bm{\phi}\)-field components to the \(\phi^{r_h}\)--\(\phi^{\Theta}\) plane gives points \(\phi(p_1), \phi(p_2), \ldots, \phi(p_n)\) along the image contour \(\Phi_i\). If these consecutive mapped points are traversed clockwise along \(\Phi_i\), the contour direction is clockwise; if traversed counterclockwise, the direction is counterclockwise. In fact contours $\Phi_i$ map the change in the components of $\phi$ as the contours $C_i$ are traversed in the \(r_h\)--\(\Theta\) plane.

As a concrete numerical example, we compute the winding number for the zero point ZP1 (\(r_h=0.500\)) from the left panel of Fig.~\ref{fig:wt} using a circular contour. For fixed parameters \(P = 0.01\), \(\lambda = 0.001\), \(Q = 0.5\), \(\tau = 11\), we construct a circular loop of radius \(0.2\) around the zero in the \((r_h, \Theta)\) plane,
\begin{equation}
\text{Circular Loop} = \left\{ \bigl( Z_i + R\cos t,\; \frac{\pi}{2} + R\sin t \bigr) \; \big| \; t \in [0, 2\pi) \right\},
\end{equation}
discretized into 12 equally spaced points. The resulting mapped points in the \(\bm{\phi}\)-plane are listed in Table~\ref{mapping}, along with the phase angle \(\Omega = \arg(\phi\)) in radians.

\begin{table}[ht]
\centering
\caption{Mapping of a circular loop around the zero at \(r_h = 0.500\) from the \((r_h, \Theta)\) plane to the \(\bm{\phi}\)-plane. The loop radius is \(0.2\). The total phase change \(\Delta \Omega_{\text{total}} = +2\pi\) confirms a winding number \(w = +1\).}
\label{mapping}
\begin{tabular}{cccc}
\toprule
Point \(i\) & \((r_h, \Theta)\) & \((\phi^{r_h}, \phi^{\Theta})\) & \(\Omega\) (radians) \\
\midrule
0 & (0.701, 1.571) & ( 0.035,  0.000) & 0.00000 \\
1 & (0.674, 1.671) & ( 0.036,  0.101) & 1.23003 \\
2 & (0.601, 1.744) & ( 0.031,  0.178) & 1.39788 \\
3 & (0.501, 1.771) & ( 0.000,  0.207) & 1.57080 \\
4 & (0.401, 1.744) & (-0.092,  0.178) & 2.04914 \\
5 & (0.328, 1.671) & (-0.257,  0.101) & 2.76720 \\
6 & (0.301, 1.571) & (-0.364,  0.000) & 3.14159 \\
7 & (0.328, 1.471) & (-0.257, -0.101) & -2.76720 \\
8 & (0.401, 1.398) & (-0.092, -0.178) & -2.04914 \\
9 & (0.501, 1.371) & ( 0.000, -0.207) & -1.57080 \\
10 & (0.601, 1.398) & ( 0.031, -0.178) & -1.39788 \\
11 & (0.674, 1.471) & ( 0.036, -0.101) & -1.23003 \\
12 & (0.701, 1.571) & ( 0.035,  0.000) & 0.00000 \\
\bottomrule
\end{tabular}
\end{table}

The winding number \(w\) is computed by summing the incremental changes in the argument of \(\bm{\phi}\) around the loop, carefully unwrapping the phase when it crosses the branch cut at \(\pm\pi\). 
The total accumulated phase is \(\Delta \Omega_{\text{total}} = 6.283185308\) radians, yielding
\[
w = \frac{\Delta \Omega_{\text{total}}}{2\pi} =\frac{\sum_i \Delta \Omega_i}{2 \pi}=\frac{ \sum_i \left(\Omega^{(i)} - \Omega^{(i-1)} \right)}{2\pi}= \frac{6.283185308}{2\pi} = 1.000000.
\]
This numerical mapping confirms that the zero at \(r_h = 0.500\) has a winding number of \(+1\), indicating a topological charge of \(+1\). The result is in agreement with the value derived using the integration method\footnote{These steps involve unwrapping: when the phase crosses from \(\pi\) to \(-\pi\) (or vice versa), we add \(2\pi\) to get the continuous change.
For example, at step 6→7, we have:
$
\Omega^{(6)} = 3.141592653 \ (\approx \pi), \quad \Omega^{(7)} = -2.767198766
$. 
The raw difference is:\\
$
\Delta \Omega_7^{\text{raw}} = -2.767198766 - 3.141592653 = -5.908791419$. Since \(|\Delta \Omega_7^{\text{raw}}| > \pi\), this indicates crossing the branch cut. Thus, we unwrap:
$
\Delta \Omega_7^{\text{unwrapped}} = -5.908791419 + 2\pi = 0.374393887
$. 
This represents the true angular change going the "long way" around the circle.}.

The left panel of Fig.~\ref{fig:con1115} shows the mapping of the contours from the left panel of Fig.~\ref{fig:wt}, and the right panel shows the mapping for the contours from the left panel of Fig.~\ref{fig:ww2}.

The topological analysis of the large-\(Q\) regime consistently reveals the Case III classification with three zero points having winding numbers \(+1\), \(-1\), and \(+1\). This structure is robust under parameter variations and is confirmed through multiple complementary approaches: direct calculation of winding numbers, visualization of vector field topology, analysis of \(r_h\)--\(\tau\) diagrams with generation/annihilation points, and mapping to the \(\bm{\phi}\)-plane.

\subsection{Topological Study for Small $Q$ Regime}
\label{tsc}

As our conventional thermodynamic analysis in Sec.~2 showed, for small values of the Maxwell charge (\(Q \lesssim 0.12\)), a Hawking-Page-type phase transition is observed. The behaviors of the free energy, temperature, and heat capacity plots confirm this. We now study the topological behavior of the system in this regime.

We begin with the parameter values \(P = 0.08\), \(Q = 0.01\), and \(\tau = 3\). Using the condition \(\partial \mathcal{F} / \partial r_h = 0\), we find two real physical horizons corresponding to topological defects (zero points) at \(r_{h1} = 0.27\) and \(r_{h2} = 1.80\). The corresponding winding numbers can be calculated by evaluating the Brouwer degree \(\eta_i = \text{sign}(J^0(\phi/x))\) at each zero. For an isolated, non-degenerate zero of a smooth vector field, the winding number \(w_i\) equals the Brouwer degree \(\eta_i\), which is determined by the sign of the Jacobian determinant \(J^0(\phi/x)\) at the zero point. This equivalence bypasses the need for explicit contour integration in Eq.~(\ref{winint}) and is established in the topological current theory of Duan \cite{Duan1984}. The Jacobian determinant \(J(r_h, \Theta)\) in the \((r_h, \Theta)\) plane is:

\begin{equation}
J(r_h, \Theta) 
= \frac{\partial \phi^{r_h}}{\partial r_h} \frac{\partial \phi^{\Theta}}{\partial \Theta} - \frac{\partial \phi^{r_h}}{\partial \Theta} \frac{\partial \phi^{\Theta}}{\partial r_h}.
\end{equation}

For our vector field construction \(\phi = (\phi^{r_h}, \phi^{\Theta})\), with \(\phi^{\Theta} = -\cot\Theta\csc\Theta\) (independent of \(r_h\)) and \(\phi^{r_h}\) independent of \(\Theta\), this simplifies to \(J = \partial^2\mathcal{F}/\partial r_h^2|_{z^{(i)}}\) at a zero point \((z^{(i)}, \pi/2)\), whose sign directly indicates thermodynamic stability. The winding numbers around these points are \(w_1 = -1\) (unstable small black hole) and \(w_2 = +1\) (stable large black hole), yielding a global topological number \(W = 0\). This is characteristic of systems exhibiting Hawking-Page-type phase transitions.

\begin{figure}[h!]
\centering
\subfloat[Vector field topology for \(Q=0.01\), \(P=0.08\), \(\tau=3\). Zero points are at \((r_h,\Theta) = (0.27,\pi/2)\) and \((1.80,\pi/2)\).\label{subfig:ww3a}]{\includegraphics[width=7cm]{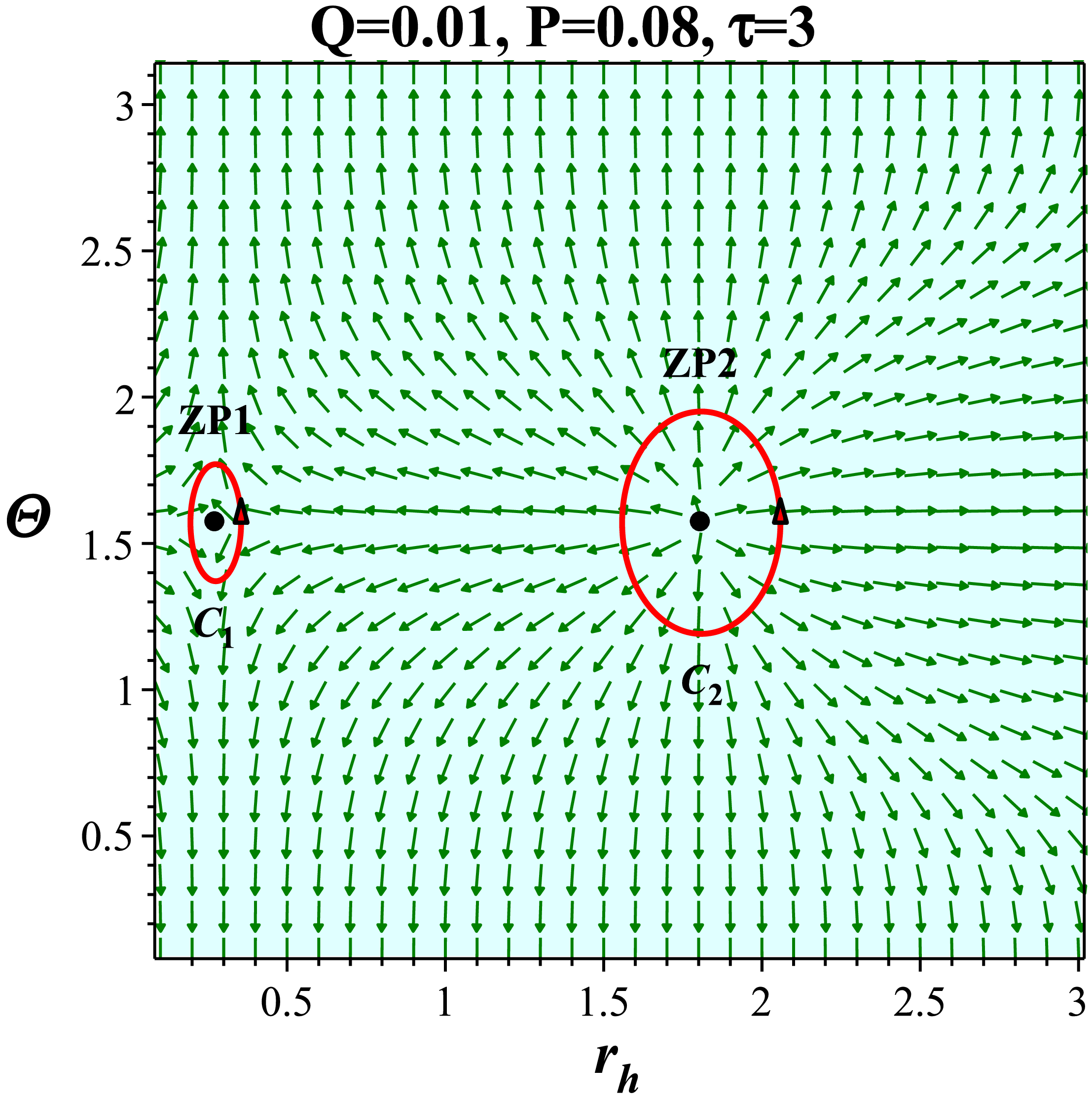}}\qquad
\subfloat[Vector field topology for \(Q=0.1\), \(P=0.01\), \(\tau=10\). Zero points are at \(r_h = 0.78\) and \(1.84\).\label{subfig:ww3b}]{\includegraphics[width=7cm]{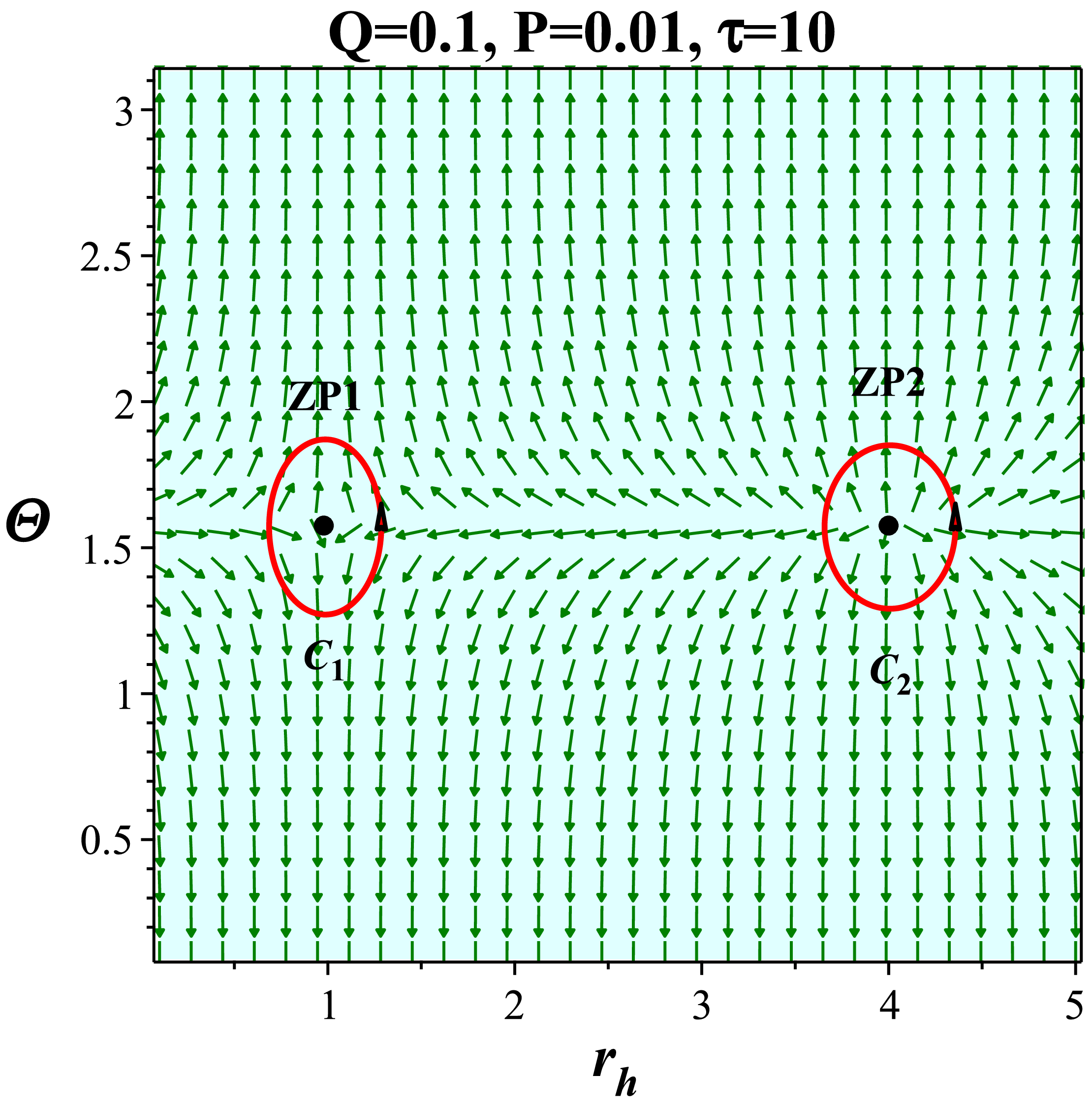}}
\caption{Vector field topology showing zero points (black dots) and integration contours (red curves). The boundary behavior on \(I_1\) and \(I_3\) indicates that the system belongs to Case IV of the topological classification.}
\label{fig:ww3}
\end{figure}

Figure~\ref{fig:ww3} shows the zero points (marked with black dots) within the \(r_h\)–\(\Theta\) plane along with the associated vector fields. In Fig.~\ref{fig:ww3}(a) for \(Q = 0.01\), \(P = 0.08\), \(\tau = 3\), the zero points are located at \((r_h, \Theta) = (0.27, \pi/2)\) and \((1.80, \pi/2)\). The red contours \(C_i\) enclose these zero points.

For comparison, Fig.~\ref{fig:ww3}(b) shows the vector field topology for \(Q = 0.1\), \(P = 0.01\), and \(\tau = 10\), where zero points occur at \(r_h = 0.78\) and \(r_h = 1.84\). The winding numbers remain \(w = (-1, +1)\), preserving the \(W = 0\) classification. The asymptotic behavior of \(\phi^{r_h}\) along the boundaries \(I_1\) and \(I_3\) confirms that the small-\(Q\) regime belongs to \textit{Case IV} in the topological classification. Consequently, \(\phi^{r_h}\) points rightward (\(\rightarrow\)) on both \(I_1\) and \(I_3\), matching the behavior of Schwarzschild-AdS black holes. This can be seen in both panels of Fig.~\ref{fig:ww3}.

\begin{figure}[H]
\centering
\subfloat[\(r_h\)–\(\tau\) diagram for \(Q=0.01\), \(P=0.08\) with \(\tau_{\text{c}} = 4.43\).\label{subfig:tt2a}]{\includegraphics[width=7cm]{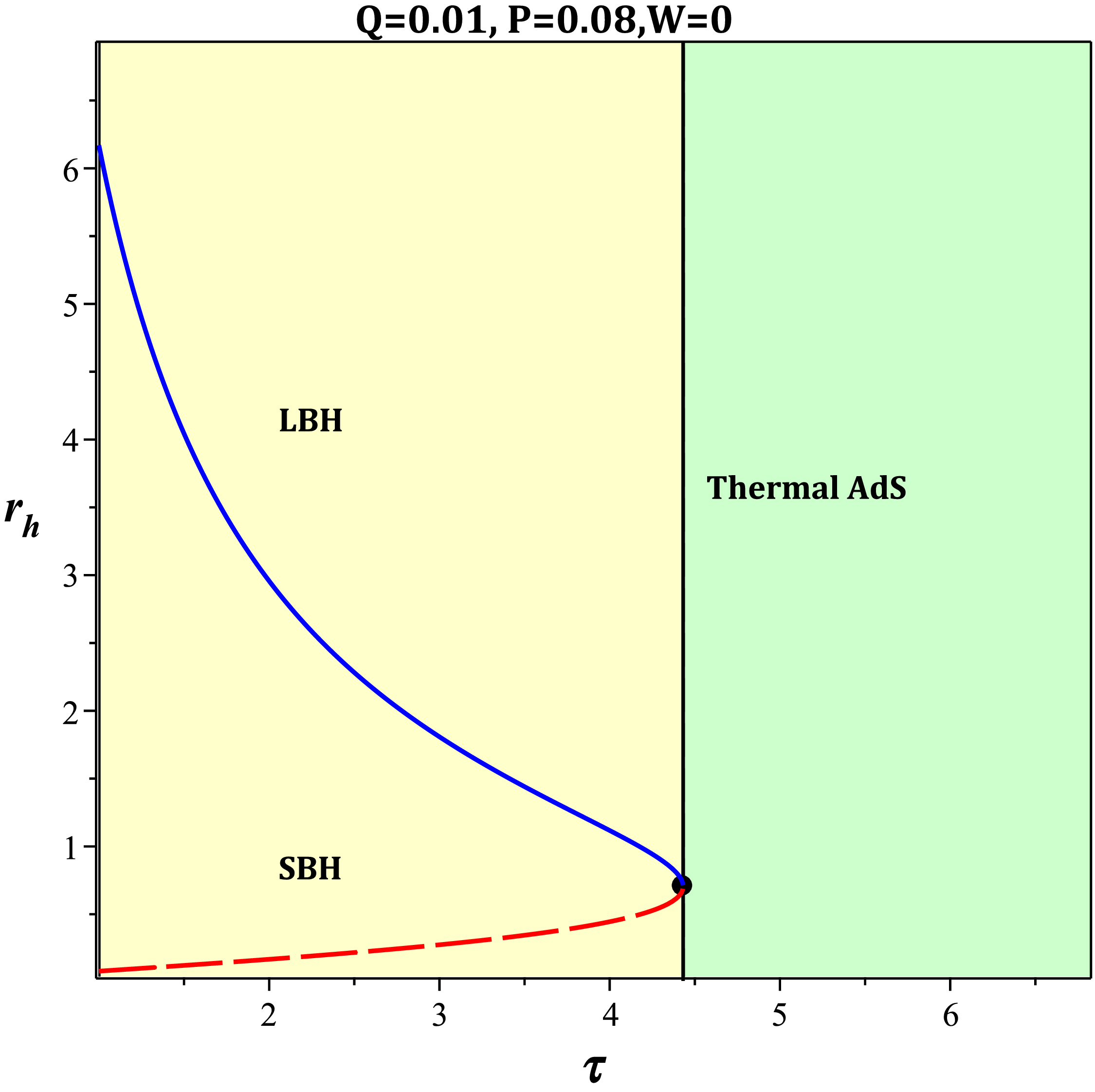}}\qquad
\subfloat[\(r_h\)–\(\tau\) diagram for \(Q=0.1\), \(P=0.01\) with \(\tau_{\text{c}} = 12.54\).\label{subfig:tt2b}]{\includegraphics[width=7cm]{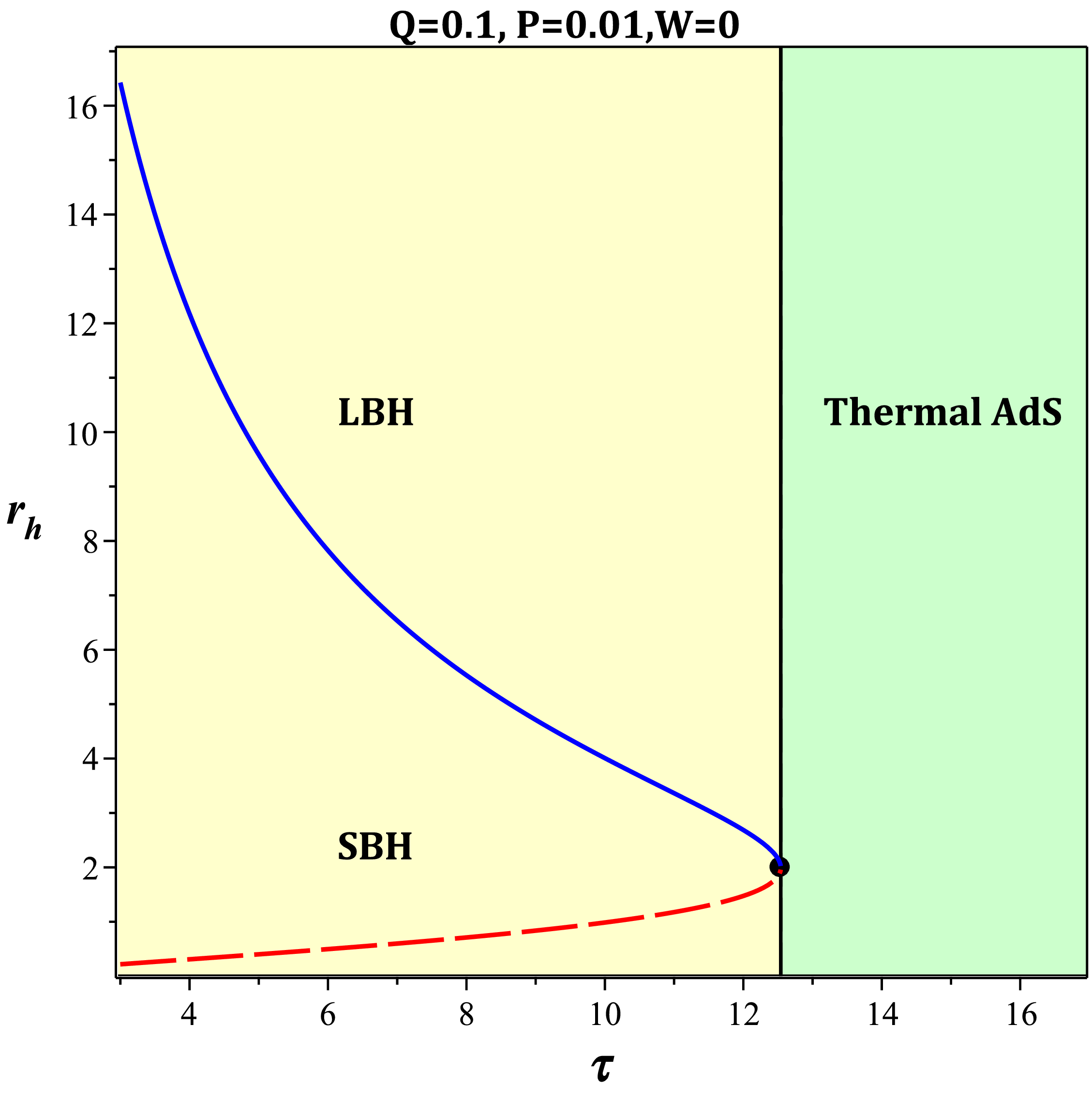}}
\caption{\(r_h\)–\(\tau\) diagrams showing two distinct black hole branches in the small-\(Q\) regime. The upper branch corresponds to the stable large black hole, while the lower branch represents the unstable small black hole.}
\label{fig:tt2}
\end{figure}

The \(r_h\)–\(\tau\) diagrams for this regime (Fig.~\ref{fig:tt2}) show two distinct black hole branches. Fig.~\ref{fig:tt2}(a) corresponds to \(Q = 0.01\), \(P = 0.08\) and displays an unstable small black hole branch (lower) and a stable large black hole branch (upper).  Unlike the large-\(Q\) regime, no swallowtail structure or generation/annihilation points appear, consistent with the absence of first-order van der Waals-type transitions; only an annihilation point is present in this regime. Fig.~\ref{fig:tt2}(b) shows similar behavior for \(Q = 0.1\), \(P = 0.01\). In each colored rectangular region, the topological invariant remains \(W = 0\). As in the usual thermodynamic analysis, we expect that above a minimum temperature two black hole states (small and large) coexist, while below it only thermal AdS space exists. In this topological picture, this translates to thermal AdS being present above a critical \(\tau_c\) and black hole states appearing below it.

\begin{figure}[H]
\centering
\subfloat[Mapping for \(Q=0.01\), \(P=0.08\), \(\tau=3\). The clockwise loop corresponds to \(w=-1\) (unstable SBH), the counterclockwise loop to \(w=+1\) (stable LBH).\label{subfig:con222a}]{\includegraphics[width=7cm]{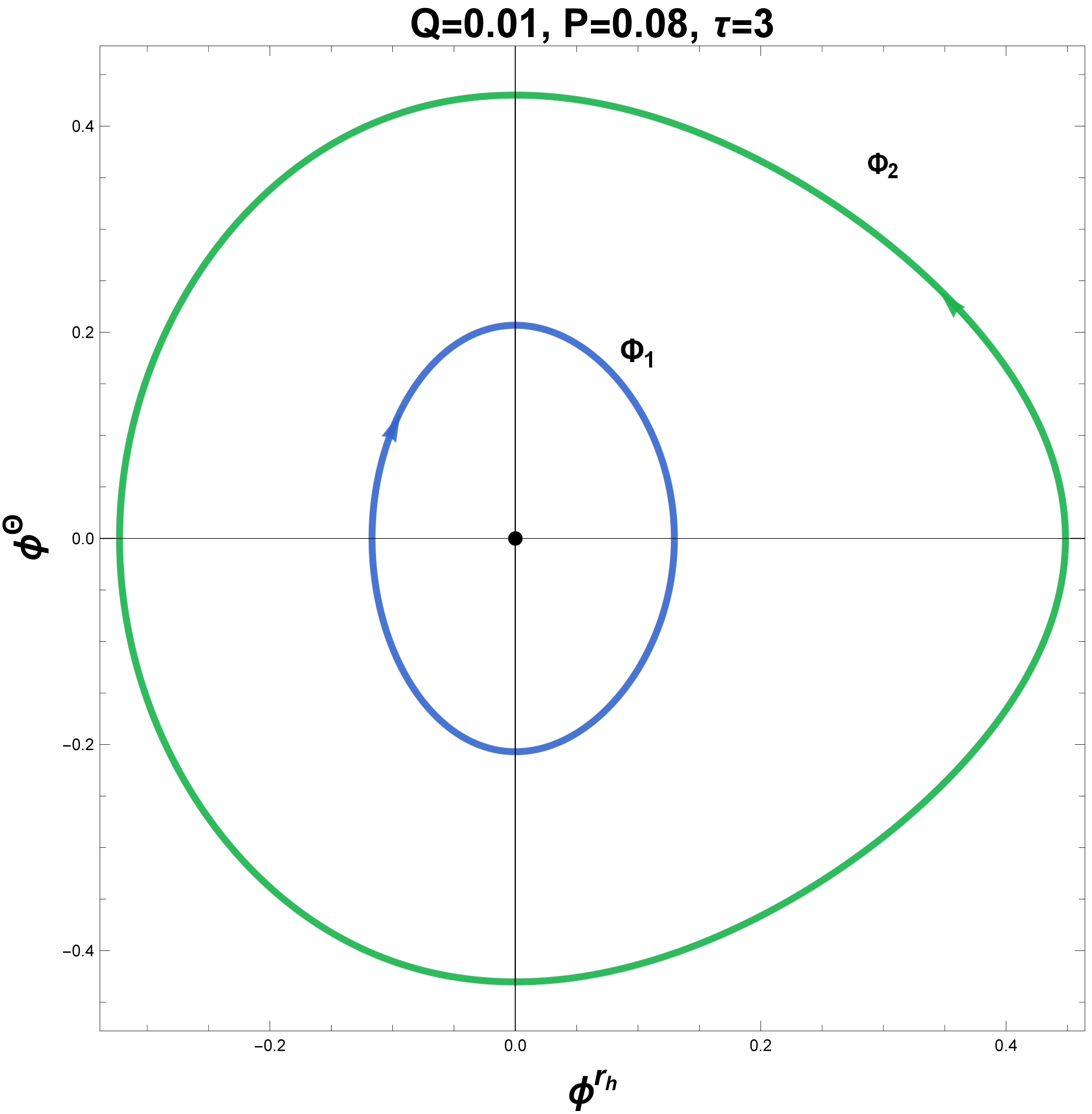}}\qquad
\subfloat[Mapping for \(Q=0.1\), \(P=0.01\), \(\tau=10\). Similar winding pattern confirms the topological classification.\label{subfig:con222b}]{\includegraphics[width=7cm]{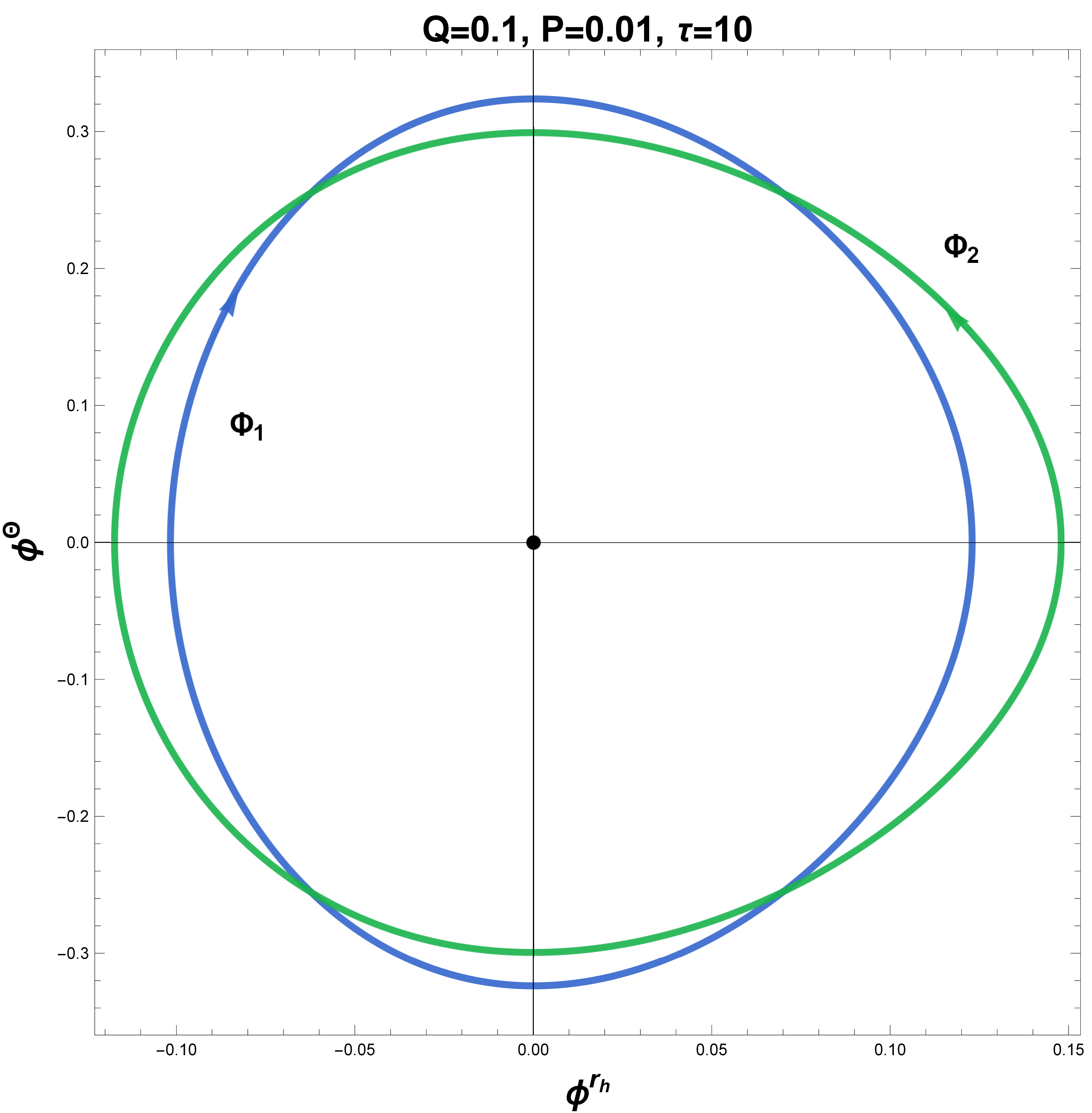}}
\caption{Mapping of contours from the \(r_h\)–\(\Theta\) plane to the \(\phi^{r_h}\)–\(\phi^{\Theta}\) plane, illustrating the winding behavior of the $\phi$-field around zero points.}
\label{fig:con222}
\end{figure}

To further verify the winding numbers, we illustrate the contours $\Phi_i$ in the \(\phi^{r_h}\)–\(\phi^{\Theta}\) plane which map the changes in the components of $\phi$ as the contours are traversed in \(r_h\)–\(\Theta\) plane. Fig.~\ref{fig:con222}(a) shows the mapping for \(Q = 0.01\), \(P = 0.08\), \(\tau = 3\). The contour around the small black hole zero maps to a clockwise loop in the \(\phi\)-plane (winding \(-1\)), while the contour around the large black hole zero maps to a counterclockwise loop (winding \(+1\)). Fig.~\ref{fig:con222}(b) shows similar behavior for \(Q = 0.1\), \(P = 0.01\), \(\tau = 10\). These diagrams confirm that the calculated winding numbers and their physical interpretation are correct.

The topological analysis of the small-\(Q\) regime consistently reveals that the system belongs to Case IV of the topological classification. This structure is robust under parameter variations and is confirmed through multiple complementary approaches: direct calculation of winding numbers, visualization of vector field topology, analysis of \(r_h\)–\(\tau\) diagrams, and mapping to the \(\phi\)-plane. The \(W = 0\) class indicates a balance between unstable and stable branches, characteristic of Hawking-Page-type transitions where thermal AdS is favored at low temperatures and large black holes dominate at high temperatures.

\section{Conclusion}
\label{sec4}

This study has employed the framework of topological thermodynamics to provide a universal classification of the phase structure of a four-dimensional AdS black hole with non-minimal Maxwell coupling. By constructing a generalized free energy and treating its critical points as topological defects in the parameter space, we have assigned a  winding number to each black hole branch and derived a global topological invariant, $W$, for the system.

Our analysis reveals a remarkable duality in the thermodynamic character of the black hole, governed by the magnitude of its Maxwell charge $Q$:
\begin{itemize}
    \item For \textbf{large charges} ($Q \gtrsim 0.12$), the system falls into the topological class $W = +1$ (Case III), characterized by the winding pattern $(+1, -1, +1)$. This topology is the definitive signature of a van der Waals-type fluid, manifesting as a first-order small/large black hole phase transition with a characteristic swallowtail structure in the free energy.
    \item For \textbf{small charges} ($Q \lesssim 0.12$), the topology shifts to the class $W = 0$ (Case IV), with the winding pattern $(-1, +1)$. This class is emblematic of a Hawking-Page transition, where thermal AdS space is globally preferred at low temperatures, yielding to a stable large black hole phase above a critical temperature, with an unstable small black hole branch acting as a separator.
\end{itemize}

\noindent The central significance of our work lies in two key insights that transcend the specifics of our model:
\begin{enumerate}
    \item \textbf{Topology as a Unifying Language:} We have demonstrated that the intricate phase behavior uncovered through conventional analysis~\cite{Sadeghi2025}, which interpolates between Hawking-Page and van der Waals types, is not a mere coincidence of parameters but is fundamentally rooted in a \textit{transition between distinct topological universality classes}. The topological number $W$ serves as a robust, model-independent order parameter that distills the complex thermodynamic landscape into its essential, global structure. This validates the physical reality of the phases from a more profound, geometric perspective.
    
    \item \textbf{Generalizing the Hawking-Page Paradigm:} A particularly consequential finding is that the \textit{non-minimal coupling $\lambda$ enables the Hawking-Page universality class ($W=0$) to persist for black holes with non-zero Maxwell charge}. In the standard Reissner-Nordstr\"om-AdS solution, this class exists only at exactly $Q=0$. Our result shows that modifications to the Einstein-Maxwell Lagrangian, such as the non-minimal coupling studied here, can \textit{stabilize this topological class against the introduction of charge}, thereby extending the realm of Hawking-Page physics. This establishes a novel link between microscopic coupling constants in the gravity action and macroscopic topological classes in thermodynamics.
\end{enumerate}

\noindent In summary, this work accomplishes more than a novel application of a topological method to a known solution. It establishes a concrete example of how topological classification can decode and predict thermodynamic behavior across different regimes of a theory. Furthermore, it identifies a new phenomenological effect in modified gravity: the stabilization of a Schwarzschild-AdS-like topological phase in charged black holes. This suggests that similar topological analyses could be used to "fingerprint" other modified gravity theories and constrain their coupling parameters based on observed or desired thermodynamic properties.

This research opens several promising avenues. The topological approach could be applied to other non-minimally coupled systems (e.g., with scalar or Yang-Mills fields) to build a taxonomy of thermodynamic classes in modified gravity. Furthermore, the observed topological transition likely has a holographic counterpart in the dual boundary conformal field theory, potentially manifesting as a change in the dominance of certain field theory phases or entanglement structures. Investigating this AdS/CFT correspondence would deepen the quantum information interpretation of black hole topological classes.

\vspace{1cm}
\noindent {\large {\bf Data Availability} } Data sharing not applicable to this article as no datasets were generated or analysed during the current study.


\end{document}